\documentclass[11pt]{article}
\pdfoutput=1

\usepackage{jheppub}

\usepackage{amsmath, amssymb} 
\usepackage{mathtools}
\usepackage{esint}
\usepackage{bm}
\usepackage{dsfont}
\usepackage{braket}
\usepackage{eqparbox}
\usepackage{enumitem}
\hypersetup{
	colorlinks,
	urlcolor=Maroon,
	linkcolor=Maroon,
	citecolor=Maroon
	}

\numberwithin{equation}{section}

\usepackage[dvipsnames,table]{xcolor}
\usepackage{soul} 
\usepackage{calc}
\usepackage{subcaption}
\usepackage{tikz}
\usepackage{tkz-euclide} 
\usetikzlibrary{arrows.meta}
\tikzset{hidden/.style = {thick, dashed}}
\usepackage{scalefnt}
\usepackage[font=small,labelfont=bf]{caption}
\usepackage{graphicx}

\usepackage[numbers, sort&compress]{natbib}
\usepackage{comment} 
\usepackage[colorinlistoftodos]{todonotes}

\usepackage{hyperref}
\hypersetup{colorlinks=true, linktoc=page, linkcolor=Maroon, citecolor=Maroon}

\newcommand{\Ad}{\mathcal{A}}

\newcommand{\s}{\sigma}

\renewcommand{\>}{\right\rangle}
\newcommand{\Tr}{\text{Tr}}
\renewcommand{\P}{\Phi}

\newcommand{\be}{\begin{equation}}
\newcommand{\ee}{\end{equation}}
\newcommand{\reef}[1]{(\ref{#1})}
\newcommand{\gap}{\text{gap}}

\usepackage{multirow}

\title{Splitting Regions and Shrinking Islands from Higher Point Constraints}

\date{\today}

\author[a]{Justin Berman,}
\author[a,b]{Henriette Elvang,}
\author[c]{Carolina Figueiredo}

\affiliation[a]{
    Leinweber Center for Theoretical Physics, Randall Laboratory of Physics\\
    University of Michigan, Ann Arbor\\
    450 Church St, Ann Arbor, MI 48109-1040, USA}
\affiliation[b]{Niels Bohr International Academy, Niels Bohr Institute \\
University of Copenhagen \\
Blegdamsvej 17, DK-2100 Copenhagen Ø, Denmark}
\affiliation[c]{Department of Physics, Jadwin Hall \\
Princeton University \\
Princeton, NJ 08540, USA}

\emailAdd{jdhb@umich.edu}
\emailAdd{elvang@umich.edu}
\emailAdd{cfigueiredo@princeton.edu}

\abstract{
We study constraints from higher-point amplitudes on $2 \to 2$ scattering in the context of effective field theory (EFT) using the perturbative numerical S-matrix bootstrap. Specifically, we investigate the class of weakly coupled EFTs with amplitudes that obey the hidden zero and split conditions that are known to hold both for $\Tr(\Phi^3)$ theory and for certain string tree amplitudes, including at 4-point the beta function. Requiring the splitting condition for the 5-point amplitude not only fixes nearly all its contact terms, but it also imposes non-linear constraints among the 4-point EFT Wilson coefficients. When included in the bootstrap, the resulting allowed region consistent with positivity is no longer convex but is restricted to a smaller non-convex  region --- which has a sharp corner near the string beta function! Assuming the absence of an infinite spin tower at the mass gap, the allowed region bifurcates into a trivial region (with states only above a chosen cutoff) and an island that continues to shrink around the string as more constraints are included in the bootstrap. The numerics indicate that in the absence of single-mass infinite spin towers the string beta function is the unique 4-point amplitude compatible with hidden zero and the 5-point splitting constraints. 
The analysis provides a prototype example for how features of higher-point amplitudes 
constrain the bootstrap of 4-point amplitudes.
}

\begin{document}

\begin{flushright}
{\tt LITP-25-08} \\
\end{flushright}
\maketitle
\flushbottom

\addtocontents{toc}{\protect\setcounter{tocdepth}{2}}

\newpage 
\section{Introduction}\label{sec:intro}

The understanding of fundamental consistency conditions such as unitarity, causality, and locality of 4-point scattering amplitudes has led to a rich program of ``bootstrapping'' $2\to2$ amplitudes  \cite{Eden:1966dnq,chew1966analytic,Veneziano:1968yb,Lovelace:1968kjy,Shapiro:1969km,Virasoro:1969me,Shapiro:1970gy,Cappelli:2012cto,Camanho:2014apa,Caron-Huot:2016icg,Paulos:2017fhb,deRham:2017avq,Caron-Huot:2020cmc,Guerrieri:2020bto,Hebbar:2020ukp,Huang:2020nqy,Arkani-Hamed:2020blm,Tolley:2020gtv,Sinha:2020win,Bellazzini:2020cot,Chiang:2021ziz,Guerrieri:2021tak,Caron-Huot:2021rmr,Bellazzini:2021oaj,He:2021eqn,Alberte:2021dnj,Henriksson:2021ymi,deRham:2021bll,Guerrieri:2021ivu,Raman:2021pkf,Chowdhury:2021ynh,CarrilloGonzalez:2022fwg,Chiang:2022ltp,Albert:2022oes,Chiang:2022jep,Haring:2022cyf,Fernandez:2022kzi,EliasMiro:2022xaa,deRham:2022gfe,Acanfora:2023axz,McPeak:2023wmq,Albert:2023jtd,Albert:2023seb,Li:2023qzs,Tourkine:2023xtu,EliasMiro:2023fqi,Berman:2023fes,Haring:2023zwu,Chiang:2023quf,Bellazzini:2023nqj,Hong:2023zgm,He:2023lyy,EliasMiro:2023fqi,Bhat:2023puy,Eckner:2024ggx,Wan:2024eto,Berman:2024eid,Guerrieri:2024jkn,Caron-Huot:2024lbf, He:2024nwd,Berman:2024wyt,Albert:2024yap,Guerrieri:2024jkn,Hillman:2024ouy,Bhat:2024agd, Beadle:2024hqg,Caron-Huot:2024tsk,Berman:2024owc,Xu:2024iao,Buoninfante:2024ibt,Gumus:2024lmj,Bhat:2025zex,Chang:2025cxc,Beadle:2025cdx,He:2025gws,Pasiecznik:2025eqc}. 
However, it is not clear {\em a priori} if a given 4-point amplitude
that passes the bootstrap constraints is also part of a physical
S-matrix with $n$-point scattering arising from a sensible quantum
field theory. 
While it remains elusive how to practically implement unitarity constraints on $n>4$ point amplitudes in order to generalize the $2\to2$ amplitude bootstrap (recent work in this direction includes \cite{Arkani-Hamed:2023jwn,Guerrieri:2024ckc,Cheung:2025krg}), progress can be made using the factorization of higher-point amplitudes into lower-point ones on their physical poles.  
Specifically, 4-point scattering processes can be constrained in non-trivial ways  via higher-point factorization when the S-matrix has certain symmetries or special kinematic properties. In this paper, we demonstrate this in a prototype example. Working in the weak coupling limit, we assume that the $n$-point amplitudes have the {\em hidden zeros and splits} recently discussed in \cite{Cachazo:2021wsz,Arkani-Hamed:2023swr,Arkani-Hamed:2024fyd}. Not only does this severely constrain the form of higher-point amplitudes, but it  gives a significant reduction of the allowed parameter space resulting from the positivity bootstrap of the 4-point amplitude. 

We briefly review the hidden zeros and splits, then outline the bootstrap setup, and follow this by a summary of our results. 

\vspace{1mm}
\noindent {\bf Hidden Zeros and Splits.}
It was shown in \cite{Arkani-Hamed:2023swr} that tree-level amplitudes of three different colored theories --- Tr$(\Phi^3)$ theory, the non-linear sigma model, and Yang-Mills theory --- share a broad class of \textit{hidden zeros}: simple loci in kinematic space where the amplitude vanishes. These were first understood in the context of Tr$(\Phi^3)$ theory since the respective amplitudes can be formulated geometrically through the canonical form of a polytope living in kinematic space, the ABHY associahedron \cite{Arkani-Hamed:2017mur}. From this viewpoint, the zeros of the amplitude correspond to limits where the polytope collapses in dimension. The geometry also suggests that the amplitude has a predictable behavior near the zero: it \textit{splits}, becoming a product of lower-point amplitudes \cite{Arkani-Hamed:2023swr,Arkani-Hamed:2024fyd}. As opposed to the ordinary factorization of tree-level amplitudes which occurs on the locus where an internal propagator goes on-shell, the splits happen for kinematics \textit{away} from poles and are not related to the exchange of physical states.

In a pure Tr$(\Phi^3)$ theory, the hidden zeros and splits can be described very simply. With  scalars in the adjoint of a non-abelian (flavor) group with generators $T^a$, the $n$-point tree amplitude can be written as a sum over color-ordered amplitudes 
\begin{align}
    \mathcal{A}_n = \sum_{\text{perms } \s}\Ad_n[1\s(2\ldots n)]\,\Tr[T^{a_1}T^{\s(a_2}\cdots T^{a_n)}]\, .
\end{align}
The color-ordered amplitude $\mathcal{A}_n[12\ldots n]$ receives contributions only from planar diagrams such that, for example, the 4-point amplitude of $\Tr(\P^3)$ theory is
\begin{align}
    \Ad_4[1234] = \Ad_4(s,u) = -\frac{g^2}{s}-\frac{g^2}{u},\,
    \label{eq:A41}
\end{align}
with $g$ the 3-point coupling.\footnote{The Mandelstam invariants are defined as
$s = -(p_1+p_2)^2$, 
$t = -(p_1+p_3)^2$, and 
$u = -(p_1+p_4)^2$
with all momenta outgoing. We use the mostly-plus signature $\eta_{\mu\nu}=\text{diag}(-1,1,\cdots,1)$ and take $d\geq 4$.} The associated hidden zero of the $4$-point amplitude \eqref{eq:A41} is then given by the locus $s+u=0 \Leftrightarrow -t=0$, for which $\Ad_4(s,-s) = 0$. Similarly, the 5-point color-ordered amplitude, 
\be
    \Ad_5[12345] = \frac{g^3}{s_{12} s_{123}} + \text{cyclic} \,,
\ee
vanishes on the  hidden zero locus $s_{13} = s_{14} = 0$ as well as on cyclic permutations thereof. 

The splits of the $5$-point amplitude occur when relaxing one of the hidden zero constraints and they take the form \cite{Arkani-Hamed:2023swr} 
\begin{equation}
\begin{split}
\mathcal{A}_5 \Big|_{s_{13}=0} = \mathcal{A}_4(s_{12},s_{15}) \times \mathcal{A}_4(s_{23},s_{34})\,, \\
\mathcal{A}_5\Big|_{s_{14}=0} = \mathcal{A}_4(s_{12},s_{15}) \times \mathcal{A}_4(s_{45},s_{34})\,,
\end{split} \, 
\label{eq:5ptSplit}
\end{equation}
and their cyclic permutations. The splits \reef{eq:5ptSplit}  imply the $5$-point hidden zero. 

Even though the hidden zeros and the splits were first understood in the field theory limit via the associahedron, both turn out to be present in string amplitudes. The 4-point string  beta function amplitude  given by\footnote{The massless scalar amplitude \reef{eq:4ptStringAmp} is obtained from the tachyon Veneziano amplitude~\cite{Veneziano:1968yb} by a kinematic shift.}
\begin{equation}
\mathcal{A}_{4}[1234] = \frac{\Gamma(-\alpha' s)\Gamma(-\alpha' u)}{\Gamma(-\alpha'(s+u))} 
\label{eq:4ptStringAmp}
\end{equation}
vanishes at $t=0$ because  $\Gamma(-\alpha'(s+u))$ has a pole at $s+u = 0$. 
Its $n$-point generalization is defined in the context of  Z-theory \cite{Huang:2016tag} and has  both hidden zeros and splits at all $n$. In particular, at five points the Z-theory string amplitude has exactly the same split factorizations \eqref{eq:5ptSplit} identified in the field theory. The low-energy expansion of \reef{eq:4ptStringAmp} gives the Tr$(\Phi^3)$ amplitude \reef{eq:A41} with $g^2 = 1/\alpha'$ at leading order and a tower of polynomial terms encoding the higher derivative $\alpha'$-corrections. 

The goal of this paper is to illustrate that properties required of higher-point amplitudes very non-trivially affect the perturbative S-matrix bootstrap of 4-point amplitudes. To this end, we focus on the example of EFTs whose amplitudes have the same hidden zeros and splits as the amplitudes of Tr$(\Phi^3)$. Any higher-derivative corrections to the 3-point interaction can be moved to higher-point via field redefinitions, so without loss of generality the 3-point vertex in the EFT is that of Tr$(\Phi^3)$ and the higher-derivative corrections take the schematic form Tr$(\partial^{2k}\Phi^n)$.

At 4-point, the hidden zero requirement that the amplitude vanishes at $t=0$ can be enforced by factoring out $-t$ from the amplitude, thus writing 
the low-energy expansion as
\begin{equation}
\mathcal{A}_4[1234] = \mathcal{A}_4(s,u) =(s+u) 
\bigg(
  -\frac{g^2}{su} + \sum_{k=0}^{\infty} \sum_{0 \le q\le  k} a_{k,q} \, s^{k-q}\, u^{q} 
\bigg)
\quad 
\text{ with } \quad a_{k,k-q} = a_{k,q} \,.
  \label{eq:WCoeffZ}
\end{equation}
The Wilson coefficients $a_{k,q}$ are in one-to-one correspondence with the effective couplings of independent linear combinations of local operators Tr$(\partial^{2(k+1)}\P^4)$, with $k$ enumerating the derivative order and $q$ being a label that distinguishes couplings of independent operators at the same derivative order. 
The condition $a_{k,q} = a_{k,k-q}$  ensures crossing symmetry 
\be
  \label{crossing}
  {\mathcal{A}}_4 (s,u) = {\mathcal{A}}_4 (u,s)  \,, 
\ee
which follows from cyclicity of the color-ordered amplitude.

An ansatz for the 5-point EFT amplitude can be constructed systematically from the 3- and 4-point EFT amplitudes by explicitly matching all the pole terms and including a complete set of cyclically invariant local contact terms with arbitrary coefficients. Imposing the hidden zero conditions on the 5-point EFT amplitude fixes many of the local contact terms, but does not constrain the 4-point amplitude. 

Subjecting the 5-point EFT amplitude to the split conditions \reef{eq:5ptSplit} fixes nearly all the contact terms coefficients, but crucially also requires non-trivial constraints among the 4-point Wilson coefficients, for example we find 
\be
 a_{2,1} = \frac{3}{2}\,a_{2, 0}-\frac{1}{2 g^2}a_{0, 0}^2 \,.
 \label{exnonlin}
\ee 
At order $k$ in the derivative expansion, there are
$\lfloor k/2\rfloor$ similar nonlinear conditions leaving only the $a_{k,0}$ coefficients unfixed. The nonlinear constraints resulting from imposing the hidden zero splits at 5-point are the new ingredient that we incorporate into the bootstrap of the 4-point EFT amplitude. 

A given UV completion of Tr$(\P^3)$ theory corresponds to a particular choice of the Wilson coefficients $a_{k,q}$ (along, of course, with the effective couplings of local operators with more than four fields). Pure Tr$(\P^3)$ is simply $a_{k,q} = 0$ for all $k,q$ while the beta function amplitude in \eqref{eq:4ptStringAmp} 
has 
\begin{equation}
\label{Venakqs}
    g^2 = 1/\alpha'
    \,, \quad
    a_{0,0} = \zeta_2\alpha'\,, \quad a_{1,0} = \zeta_3\alpha'^2\,, \quad a_{2,0} = \zeta_4\alpha'^3\,, \quad a_{2,1} = \frac{1}{4}\zeta_4 \alpha'^3\, , \ldots
\end{equation}
where $\zeta_n$ is the Riemann Zeta function. 
The Wilson coefficients \reef{Venakqs} indeed satisfy the nonlinear constraint \reef{exnonlin}.

\vspace{1mm}
\noindent {\bf Bootstrap.}
In the S-matrix bootstrap, the low-energy Wilson coefficients are connected to the UV physics through dispersive integrals over the spectral density. In order to derive such dispersion relations, we assume that the amplitude is polynomially bounded in the Regge limit of large $|s|$ and fixed $u<0$ (or fixed $t<0$). 
Concretely, we assume an improved Regge behavior 
\be
  \lim_{|s| \to \infty} \frac{\mathcal{A}_4(s,u)}{s} =0 \,,~~~~~
  \lim_{|s| \to \infty} \frac{\mathcal{A}_4(s,-s-t)}{s} =0
  \label{eq:bounduandtfixed}
\ee
for $u$ and $t$ fixed, respectively. This behavior is not generic; it is a stronger assumption than the standard Froissart-Martin bound, where instead of a single power of $s$ in the denominator in \eqref{eq:bounduandtfixed}, we would have $s^2$. In Appendix \ref{app:2sdr}, we discuss what happens with this harder Regge behavior, but otherwise we shall assume \eqref{eq:bounduandtfixed} since that allows us to give the clearest illustration of the impact of the nonlinear constraints from higher points on the bootstrap bounds. 

We also assume that for physical kinematics (when $s$ and $u$ are both real and satisfy $s \geq -u >0$), the amplitude can be expanded in terms of  partial waves
\begin{equation}
\label{eq:pwexp}
  \mathcal{A}_4(s,u) = \sum_{j=0}^\infty n_j^{(d)} a_{j}(s) \mathcal{P}_j^{(d)} \left(1 + \frac{2u}{s} \right),
\end{equation}
where $\mathcal{P}_j^{(d)}$ are the $d$-dimensional Gegenbauer polynomials, $a_j(s)$ is the partial wave from spin $j$, and $n_j^{(d)}$ is a normalization factor given by 
\begin{equation}
    n_j^{(d)}=\frac{(4\pi)^{d/2}(2j+d-3)\Gamma(j+d-3)}{\pi \Gamma(\frac{d-2}{2})\Gamma(j+1)}\,.
\end{equation}
The imaginary part of  $a_j(s)$ is related to the respective spectral density
\begin{equation}
\label{eq:specden}
    \rho_j(s) = \frac{1}{\pi} n_j^{(d)} s^{d-4/2} \text{Im}\, a_j(s)\,.
\end{equation}
For the case of weakly-coupled theories, as assumed here, unitarity implies the \textit{positivity} of the spectral density $\rho_j(s)$ for physical kinematics. The derivation of the dispersion relations also relies on assuming analyticity of $\mathcal{A}_4$ in the complex $s$-plane away for the real $s$-axis as well as on the  existence of a mass gap $M_{\gap}$ such that the spectral density only has support for real $s \ge M_{\gap}^2$. The scale of $M_{\gap}$ can be used to make every Wilson coefficient dimensionless, so the bounds that result from the dispersive representations are always in units of $M_{\gap}$.

From the dispersive representations, two-sided bounds can be derived on ratios of Wilson coefficients, such as $a_{k,q}/a_{0,0}$ for all $k\ge1$. Values of the Wilson coefficients that lie outside these bounds are strictly ruled out under the given assumptions and hence any viable amplitude has to have values of the $a_{k,q}$ such that their ratios lie in the allowed region (sometimes called the ``EFT-hedron") within the bootstrap bounds.  
Since unitarity is imposed merely as positivity, any  positive linear combination of  two 4-point amplitudes with $\rho_j(s) \ge 0$ also has positive spectral density, hence in absence of any additional constraints the general allowed region is necessarily convex. 

Despite this freedom to sum amplitudes at 4-point, requiring consistent factorization of amplitudes from higher point to lower point breaks this principle. The sum of two $n$-point amplitudes which each individually can be factorized into lower points would in general not consistently factorize on all its poles. We should therefore not expect that the ability to sum amplitudes would generalize once we take into account higher point information. Indeed, this is precisely what we observe: when we impose consistent factorization together with the splits summed amplitudes are no longer valid and the allowed region is not convex.

\vspace{1mm}
\noindent {\bf Results: Bootstrap with Hidden Zero and Splits.}
A key motivation for this paper is that when amplitudes are required to satisfy additional properties or symmetries, it may no longer be the case that their linear combinations also preserve those properties. This is in fact the case for the amplitudes satisfying the hidden zeros and splits, e.g.~\reef{eq:5ptSplit}, since --- as we have seen --- the splits require the $a_{k,q}$'s to satisfy  nonlinear, nonconvex constraints such as \reef{exnonlin}. Bootstrapping EFT amplitudes subject to such nonlinear constraints, 
we may expect the general convex allowed region to be reduced to a smaller region that need not be convex. We show that this expectation is realized in the case of the splits. The {\em non-convexity} is illustrated in Figure \ref{fig:genbds}, where the new non-convex allowed region has two sharp corners: one for the beta function amplitude and the other for an infinite spin tower amplitude. 

With convexity gone, simple spectral assumptions can even break connectivity of the allowed region. We show that for  amplitudes compatible with the 5-point splits \reef{eq:5ptSplit}, the allowed region 
{\em bifurcates} into two separate parts when it is assumed that there are only a finite number of spin states at the mass gap $M_{\gap}$ and no other states for $M_{\gap}^2 < s < \mu_c M_{\gap}^2$ given a choice of cutoff $\mu_c > 1$. One region corresponds to theories with no states at $M_{\gap}^2$ and another one contains the string amplitude. As we increase the number of split non-linear constraints imposed (see Table \ref{tab:Constraintsk6}), the size of the island around the string decreases dramatically! Plots of the bifurcated regions and string islands for $d=10$ and cutoff $\mu_c=2$ as well as for $d=4$ and $\mu_c=1.2$ are shown in 
Figures \ref{fig:Zoom10d} and \ref{fig:Zoom4d}, respectively. In the former case, the allowed range of the  Wilson coefficients $a_{1,0}/a_{0,0}$, $a_{3,0}/a_{0,0}$, and the linear combination $(3a_{2,0}-2a_{2,1})/a_{0,0}$ are restricted to lie within about $10^{-5}$ of the string values. 
These numerical results indicate that {\em in the absence of an infinite tower of states at $M_\text{gap}$, the string beta function \reef{eq:4ptStringAmp} is the unique unitary 4-point amplitude compatible with the hidden zero splits \reef{eq:5ptSplit}.}

It is interesting to note that 
on top of the zeros and splits identified in the field theory limit, the string amplitudes \cite{Huang:2016tag} satisfy an infinite class of split factorizations that do not survive the field theory limit \cite{Arkani-Hamed:2023swr}. For example, the beta function amplitude  \eqref{eq:4ptStringAmp} also vanishes for  $s+u = N/\alpha^\prime$, for $N\in \mathbb{Z}^+$. It is very plausible that these $n$-point string amplitudes are the \textit{only} functions satisfying this infinite class of constraints. However, it is much less trivial if the same holds when we only ask for the zeros/splits that survive in the field theory limit, $i.e.$ those  we can discover without knowing about strings {\em a  priori}. Our analysis offers evidence that, up to cases with infinite spin towers, these string amplitudes are unique in this sense.

\vspace{1mm}
\noindent {\bf Outline.}
In Section \ref{sec:HZ}, we explain how to parametrize the $5$-point amplitude in terms of 
its pole terms (that depend on the  4-point Wilson coefficients $a_{k,q}$) and purely local terms with arbitrary coefficients. 
We use this representation, together with the constraints in \eqref{eq:5ptSplit}, to derive the nonlinear constraints among the  $a_{k,q}$ coefficients. In the process most of the coefficients for the 5-point local terms are fixed in terms of the 4-point coefficients.
Imposing splitting conditions on  the 6-point EFT amplitude does not further restrict the 4-point coefficients $a_{k,q}$, but they fix additional contact terms in the 5-point amplitude. Details of how the splitting conditions fix local terms in the 5- and 6-point EFT amplitudes are given in Appendix \ref{app:LocalTerms}. 

In Section \ref{sec:Bootstrap}, we present the dispersive representation of the 4-point Wilson coefficients $a_{k,q}$ as well as the null constraints coming from the fixed-$u$ and $t$ dispersion relations. We explain how we impose (subsets of) the nonlinear relations among the $a_{k,q}$ as constraints on the spectral density. The crucial aspect is that in order to implement the split constraints in the numerical linear solver SDPB \cite{Simmons-Duffin:2015qma}, we need to first linearize them. 
We discuss the approach in Section \ref{s:NonLin} and provide a more extensive explanation of the procedure in Appendix \ref{app:nonlin}. 
In Section \ref{s:basicbounds}, we present the results of the bootstrap and show that the allowed region can now be non-convex. 

In Section \ref{sec:spectrum}, we consider an additional assumption about the spectrum: we require that there is a \textit{finite} number of states at the mass gap, $M_{\gap}^2$, and that all remaining states are above some cutoff $\mu_c M_{\gap}^2$, with $\mu_c > 1$. The allowed region then bifurcates into a scaled-down version of the general non-convex region and an island around the string. The island shrinks as null constraints from an increasing number of higher-derivative terms are included in the bootstrap and we use this to determine numerical bounds on the lowest Wilson coefficients. 

We conclude with outlook and discussion of the results in Section \ref{sec:outlook}, including comparison of the split constraints to other contexts of nonlinear constraints of Wilson coefficients, comparison of the size of the string islands with those previously obtained using string monodromy relations, and result of relaxing the Regge boundedness assumption \reef{eq:bounduandtfixed} (with additional details in 
Appendix \ref{app:2sdr}.)

\section{Hidden Zeros and Splits for scalar EFTs}
\label{sec:HZ}

In this section we derive the constraints on the 4-point Wilson coefficients coming from the 5-point split factorizations  \eqref{eq:5ptSplit}. 

\subsection{Leading Order}
\label{s:HZreview}
Since we are considering scalar scattering, the $n$-point tree amplitude $\mathcal{A}_n$ is simply a function of the dot products of the external momenta, $p_i\cdot p_j$, on the support of momentum conservation, $\sum_i p_i^\mu =0$. In order to trivialize the constraints coming from momentum conservation on the $p_i \cdot p_j$'s, we write the most general ansatz for $\mathcal{A}_5$ in terms of the kinematic basis given by the Mandelstam invariants $X_{i,j} = (p_i + p_{i+1} + \cdots + p_{j-1})^2$. By momentum conservation, $X_{j,i}=X_{i,j}$. These variables are usually called the \textit{planar variables}, as they correspond to the Mandelstam invariants appearing as poles in color-ordered amplitudes (in the standard ordering). The $X_{i,j}$'s are also naturally associated with the length-squared of the chords of the \textit{momentum polygon} -- the polygon we obtain by associating the $n$ momenta of the external states to the sides of an $n$-gon, following the ordering the amplitude is defined on. It is possible to show that for an $n$-point amplitude, the corresponding $n$-gon has precisely $n(n-3)/2$ chords which define the same number of $X_{i,j}$, precisely matching the dimensionality of kinematic space.\footnote{Of course for general $n$ and finite $d$, there are Gram determinant constraints between the $X_{i,j}$ that make them not independent. However, for our analysis we will be interested in the case of $n=5$ and we take $d\geq 4$ in which case there are no Gram determinant constraints.}

In terms of the planar variables, the 4- and 5-point $\Tr(\P^3)$ amplitudes can be written
\begin{align}
    \label{eq:A4}
    \Ad_4[1234] &= \Ad_4(X_{1,3},X_{2,4}) = 
    -g^2 \left(\frac{1}{X_{1,3}}+\frac{1}{X_{2,4}} \right)\,,
    \\
    \label{eq:A5}
    \Ad_5[12345] &= 
    g^3 
    \left( \frac{1}{X_{1,3}X_{1,4}}+\frac{1}{X_{2,4}X_{2,5}}+\frac{1}{X_{3,5}X_{1,3}}+\frac{1}{X_{1,4}X_{2,4}}+\frac{1}{X_{2,5}X_{3,5}} \right)\,.
\end{align}
The hidden zero locus, $s_{13} = 0$ and $s_{14}=0$, translates to the conditions
\be
  \label{5ptHZlocus}
  X_{1,4} = X_{2,4}+X_{1,3}
  ~~~\text{and}~~~
  X_{2,4} = X_{1,4}+X_{2,5} \,. 
\ee
On each of these loci, the 5-point amplitude splits as follows:
\begin{eqnarray}
&g \mathcal{A}_5\big(X_{1,4}=X_{2,4}+X_{1,3}\big)= \mathcal{A}_4\big(X_{1,3},X_{2,5}\big) \times \mathcal{A}_4\big(X_{2,4},X_{3,5}\big)\,,
\label{eq:5ptSplitA}
\\
&g \mathcal{A}_5\big(X_{2,4}=X_{1,4}+X_{2,5}\big) =
\mathcal{A}_4\big(X_{1,3},X_{2,5}\big) \times \mathcal{A}_4\big(X_{1,4},X_{3,5}\big)\,,
\label{eq:5ptSplitB}
\end{eqnarray}
where the $g$ factor multiplying $ \mathcal{A}_5 $ ensures the correct units. Note that the splitting relations \reef{eq:5ptSplitA}-\reef{eq:5ptSplitB} imply the hidden zero conditions on the locus \reef{5ptHZlocus}.

Next, we construct the most general 4- and 5-point amplitudes in adjoint scalar EFTs whose leading interaction is $\Tr(\P^3)$. We then subject these amplitudes to the hidden zero and splitting constraints  
\reef{eq:5ptSplitA}-\reef{eq:5ptSplitB}.

\subsection{EFT Amplitudes and Nonlinear Constraints}
\label{s:HZEFT}

At 4-point, the most general EFT corrections can be encoded in terms of $s \leftrightarrow u$ symmetric polynomials terms in $\mathcal{A}_4$. The hidden zero condition that the amplitude vanishes at $t=0$ can be imposed by simply factoring out $s+u$: 
\be
  \label{AtildeDef}
  \mathcal{A}_4(s,u)  = (s+u) \tilde{\mathcal{A}}_4(s,u) 
   =(s+u)
   \bigg( 
     -\frac{g^2}{su} + \tilde{\mathcal{A}}_4^{\text{C}}(s,u)
   \bigg)\,,
\ee
where the contact terms in $\mathcal{A}_4$ are 
\be
  \mathcal{A} _4^{\text{C}}(s,u)
  = (s+u)
    \tilde{\mathcal{A}}_4^{\text{C}}(s,u)
    =(s+u)\sum_{k=0}^{\infty} \sum_{0 \le q\le  k} a_{k,q} \, s^{k-q}\, u^{q}\,
    \label{A4Contact}
\ee
and crossing symmetry 
$\mathcal{A}_4(s,u)=\mathcal{A}_4(u,s)$
requires 
$a_{k,k-q} = a_{k,q}$.

To construct the most general ansatz for the 5-point EFT amplitude, we divide the low-energy expansion of $\mathcal{A}_5$ into three different parts,
\be
\label{5ptansatz}
\mathcal{A}_5 = \mathcal{A}_5^{(1)}+\mathcal{A}_5^{(2)} +  \mathcal{A}_5^{\text{C}}\,,
\ee
which are
\begin{enumerate}
    \item The part with two poles, corresponding to two factorization channels and three cubic vertices
    \begin{equation}
       \mathcal{A}_5^{(1)} = \begin{gathered}
    \begin{tikzpicture}[line width=0.7,scale=0.5,baseline={([yshift=-0.5ex]current bounding box.center)}]
        \coordinate (p1) at (-1,0);
        \coordinate (p2) at (0,0);
        \coordinate (p3) at (1,0);
        \coordinate (p1b) at (-2,-1);
        \coordinate (p1t) at (-2,1);
        \coordinate (p2t) at (0,1);
        \coordinate (p3t) at (2,1);
        \coordinate (p3b) at (2,-1);

        \draw[] (p1) -- (p2) -- (p3);
        \draw[] (p1t) -- (p1);
        \draw[] (p1b) -- (p1);
        \draw[] (p2t) -- (p2);
        \draw[] (p3t) -- (p3);
         \draw[] (p3b) -- (p3);
    \node[scale=0.8,xshift=-6,yshift=6] at (p1t) {$2$};
    \node[scale=0.8,xshift=-6,yshift=-6] at (p1b) {$1$};
    \node[scale=0.8,xshift=0,yshift=7] at (p2t) {$3$};
    \node[scale=0.8,xshift=6,yshift=6] at (p3t) {$4$};
    \node[scale=0.8,xshift=6,yshift=-6] at (p3b) {$5$};
    \end{tikzpicture}
    ~~+  ~~\text{cyclic}
\end{gathered} \quad 
= \quad \frac{g^3}{X_{1,3} X_{1,4}} + \text{cyclic}\,.
    \end{equation}
    Thus $\mathcal{A}_5^{(1)}$ is simply the leading order 5-point amplitude \reef{eq:A5}.
    \item The part with a single 2-particle channel, corresponding to processes with a cubic and a quartic interaction
    \begin{equation}
       \mathcal{A}_5^{(2)} = \begin{gathered}
    \begin{tikzpicture}[line width=0.7,scale=0.5,baseline={([yshift=-0.5ex]current bounding box.center)}]
        \coordinate (p1) at (-1,0);
        \coordinate (p2) at (0.2,0);
        \coordinate (p1b) at (-2,-1);
        \coordinate (p1t) at (-2,1);
        \coordinate (p2t) at (1.2,1);
        \coordinate (p2r) at (1.2,0);
        \coordinate (p2b) at (1.2,-1);

        \draw[] (p1) -- (p2);
        \draw[] (p1t) -- (p1);
        \draw[] (p1b) -- (p1);
        \draw[] (p2t) -- (p2);
        \draw[] (p2b) -- (p2);
         \draw[] (p2r) -- (p2);
    \node[scale=0.8,xshift=-6,yshift=6] at (p1t) {$2$};
    \node[scale=0.8,xshift=-6,yshift=-6] at (p1b) {$1$};
    \node[scale=0.8,xshift=6,yshift=6] at (p2t) {$3$};
    \node[scale=0.8,xshift=6,yshift=0] at (p2r) {$4$};
    \node[scale=0.8,xshift=6,yshift=-6] at (p2b) {$5$};
    \end{tikzpicture}
    ~~~~+  ~~\text{cyclic}
\end{gathered} \quad 
= \quad-  g\frac{1}{X_{1,3}}
\,\mathcal{A}_4^{\text{C}}(X_{1,4},X_{3,5}) + \text{cyclic}\,.
\label{A5singlepole}
    \end{equation}
    Here $ \mathcal{A}_4^{\text{C}}$ denotes the contact terms of the 4-point amplitude \reef{A4Contact}.
    \item The part corresponding to $5$-point contact/local terms. This can be written as a general expansion in cyclically symmetric polynomials of the $X_{i,j}$'s, giving:
    \begin{equation}
    \begin{aligned}
       \mathcal{A}_5^{\text{C}} =  & \, w_{0,0} + w_{1,0} \big[ X_{1,3} + X_{2,4} + X_{3,5} + X_{1,4} +X_{2,5}\big] \\
       &+ w_{2,0}\big[X_{1,3}^2 + \text{cyclic}\big] + w_{2,1}\big[X_{1,3}X_{1,4} 
       + \text{cyclic}\big]  
       +w_{2,2}\big[X_{1,3}X_{2,4} +  \text{cyclic}\big] \\
       & +w_{3,0} \big[X_{1,3}^3 + \text{cyclic}\big] + \ldots\,.
    \end{aligned}
    \label{eq:5ptLoc}
    \end{equation}
    The $w_{i,j}$ are essentially\footnote{The ansatz \reef{5ptansatz} has the correct pole residues, however, an ambiguity arises from the ordering of the arguments in the contact terms: on-shell, $\mathcal{A}_4^{\text{C}}$ is crossing symmetric, but not off the pole. For example, for the first term in \reef{A5singlepole}, we have $\mathcal{A}_4^{\text{C}}(X_{1,4},X_{3,5})=\mathcal{A}_4^{\text{C}}(X_{3,5},X_{1,4}) + O(X_{1,3})$ for $X_{1,3} \ne 0$. However, this ambiguity affects to only local terms. Therefore \reef{A5singlepole} correctly captures the residues of each pole term and the ambiguity is compensated  by the general coefficients $w_{i,j}$ in the contact term ansatz \reef{eq:5ptLoc}. (For that reason, the $w_{i,j}$'s may not directly reflect the Wilson coefficients of the 5-field higher derivative terms in the EFT Lagrangian; this is because they can also be affected by field redefinitions.)} the $5$-point Wilson coefficients, with $i$ labelling the degree of the polynomial, and $j$ labels the different cyclic polynomials at order $\mathcal{O}(X^i)$. 
\end{enumerate}

We now impose the splitting conditions \reef{eq:5ptSplitA}-\reef{eq:5ptSplitB} order-by-order in the low-energy expansion. At leading order, $\mathcal{O}(X^{-2})$, 
the splitting conditions are satisfied because this is just the Tr$(\Phi^3)$ theory. To illustrate the procedure at the next orders in $X$, we focus on \reef{eq:5ptSplitA}. 
On the right hand side of \reef{eq:5ptSplitA}, the $\mathcal{O}(X^{-1})$ terms are
\be
\label{A4xA4poles}
\begin{split}
&\mathcal{A}_4\big(X_{1,3},X_{2,5}\big) \times \mathcal{A}_4\big(X_{2,4},X_{3,5}\big)
\\
& \hspace{1.5cm}
\supset
-\bigg(
\frac{g^2}{X_{1,3}} + \frac{g^2}{X_{2,5}}
\bigg)
\mathcal{A}_4^\text{C}\big(X_{2,4},X_{3,5}\big)
- 
\bigg(
\frac{g^2}{X_{2,4}} + \frac{g^2}{X_{3,5}}
\bigg)
\mathcal{A}_4^\text{C}\big(X_{1,3},X_{2,5}\big)
\end{split}
\ee
while on the left hand side, the $\mathcal{O}(X^{-1})$ contributions arise from the five terms \reef{A5singlepole} which give 
\begin{equation}
\begin{aligned}
g \mathcal{A}_5\big(X_{1,4} 
=
X_{1,3}+X_{2,4}\big)
\supset &  ~-\frac{g^2}{X_{3,5}} \mathcal{A}_4^{\text{C}} (X_{1,3},X_{2,5})
- \frac{g^2}{X_{2,5}} \mathcal{A}_4^{\text{C}}(X_{2,4},X_{3,5}) 
\\
&
-\frac{g^2}{X_{2,4}} \mathcal{A}^{\text{C}}_{4}(X_{1,3}+X_{2,4},X_{2,5})  
-
\frac{g^2}{X_{1,3}} \mathcal{A}_4^{\text{C}}(X_{1,3}+X_{2,4},X_{3,5}) 
\\
&-\frac{g^2}{X_{1,3}+X_{2,4}} \mathcal{A}_4^{\text{C}}(X_{1,3},X_{2,4}) \, .
\end{aligned}
\label{eq:PolesFact5}
\end{equation}
The two terms in the first line of 
\reef{eq:PolesFact5} directly cancel against equivalent terms on the right hand side of \reef{eq:5ptSplitA}.  
The two terms in the second line of \reef{eq:PolesFact5} have residues at $X_{1,3} = 0$ and $X_{2,4} = 0$, respectively, that cancel against the two matching terms in \reef{A4xA4poles} and leave behind terms of $\mathcal{O}(X^0)$ and higher. The term in the last line of  
\reef{eq:PolesFact5} is itself $\mathcal{O}(X^0)$, since by \reef{A4Contact} it is simply equal to 
$g^2 \tilde{\mathcal{A}}_4^{\text{C}}(X_{1,3},X_{2,4})$. Thus, all pole terms in \reef{eq:5ptSplitA} cancel.

We now proceed to match the polynomial contributions. 
Moving all terms to one side of the equation, \reef{eq:5ptSplitA} requires the vanishing of
\begin{align}
   \label{5ptmaster}
   0 \,=~& g^2\sum_{n=1}^\infty \frac{X_{2,4}^{n-1}}{n!} \frac{\partial^n \mathcal{A}_4^{\text{C}}(X_{1,3},X_{2,5})}{\partial X_{1,3}^n} 
   +
   g^2\sum_{n=1}^\infty
   \frac{X_{1,3}^{n-1}}{n!} \frac{\partial^n \mathcal{A}_4^{\text{C}}(X_{2,4},X_{3,5})}{\partial X_{2,4}^n}
   \\
   &
   + g^2\tilde{\mathcal{A}}_4^{\text{C}}(X_{1,3},X_{2,4})
   -g\mathcal{A}_5^{\text{C}}\big(X_{2,4}=X_{1,4}+X_{2,5}\big)
   +\mathcal{A}_4^{\text{C}}
   \big( X_{1,3},X_{2,5}\big)
   \mathcal{A}_4^{\text{C}}
   \big( X_{2,4},X_{3,5}\big)\,.
   \nonumber
\end{align}
Equation \reef{5ptmaster} is now solved order-by-order in the $X$-expansion. 
At the lowest orders, we  find
\begin{equation}
   \mathcal{O}(X^0): \quad  w_{0,0} = 3 g a_{0,0}\,, \quad \quad \mathcal{O}(X^1): \quad w_{1,0} = 2 g a_{1,0} \,,
\end{equation}
so this fixes the all $5$-point Wilson coefficients at the two lowest orders. 

The next order, $\mathcal{O}(X^2)$, is more interesting. Including  select $\mathcal{O}(X^2)$ mononomials in the expansion of \reef{5ptmaster}, we have
\be
  \begin{split}
  0 =&  X_{2,5}^2 
  \big[g^2 a_{2,0} + g^2 a_{2,1} 
  - g w_{2,0}\big]
  + 
  X_{1,3} X_{2,5} 
  \big[2g^2 a_{2,0} + 2g^2 a_{2,1} - 2g w_{2,2}\big]
  \\
  &~
  +
  X_{2,5} X_{3,5} 
  \big[a_{0,0}^2 - g w_{2,1}\big]
  + 
  X_{1,3}^2 
  \big[5g^2 a_{2,0} - 2 g w_{2,0}-  g w_{2,1}\big] + \ldots
  \end{split}
\ee
The vanishing of the first three coefficients fixes the three 5-point Wilson coefficients $w_{2,j}$. Using those,  the coefficient of $X_{1,3}^2$ vanishes only when the 4-point Wilson coefficients $a_{2,0}$, $a_{2,1}$, and $a_{0,0}$ satisfy the nonlinear relationship \reef{exnonlin} we presented in the Introduction. 
With these relations, all other $\mathcal{O}(X^2)$ terms on the right hand side of \reef{5ptmaster} vanish. Thus, at this order we have determined  
\begin{equation}
    w_{2,0} = w_{2,2} 
    =  \frac{5}{2}\, g \,a_{2,0} -\frac{1}{2g}a_{0,0}^2\,, 
    \quad w_{2,1} = \frac{a_{0,0}^2}{g}\,, 
    \quad a_{2,1} = \frac{3}{2}\,a_{2, 0}-\frac{1}{2 g^2}a_{0, 0}^2\,.
\end{equation}
It may seem surprising that constraints on the 4-point coefficients $a_{k,q}$ arise from the polynomial terms in \reef{5ptmaster} when the 5-point ansatz \reef{5ptansatz} includes all cyclic polynomial terms with free coefficients $w_{i,j}$ by construction. However, the product $\mathcal{A}_4 \times \mathcal{A}_4$ on the RHS of the splitting formula \reef{eq:5ptSplitA} and the particular kinematic limit taken breaks the cyclic symmetry. Thus, there is a larger class of local polynomials that can be produced in the special kinematic limit from the terms in $\mathcal{A}_4 \times \mathcal{A}_4$ and the polynomials terms from $\mathcal{A}_5^{(2)}$ than from the particular kinematic limit of the cyclic polynomials in $\mathcal{A}_5^{\text{C}}$. That is why fixing $w_{i,j}$ is not sufficient for solving the splitting constraints and how the constraints on the $a_{k,q}$ arise.

As we extend the analysis to higher orders in the momentum expansion, we have to impose both 5-point splits, \eqref{eq:5ptSplitA} and \eqref{eq:5ptSplitB}, 
 to derive all the constraints.\footnote{The remaining 5-point splits are related to these two by cyclic permutations, so they are  automatically satisfied because our 5-point ansatz is cyclically invariant.} 
 Working up to $\mathcal{O}(X^{14})$, we find that all $a_{k,q}$ are fixed except the $a_{k,0}$ which remain as the only free parameters in the low-energy ansatz of the 4-point EFT amplitude. The nonlinear constraints for 
$k=2,3,4,5,6$ are listed explicitly in Table \ref{tab:Constraintsk6}. If $g^2$ is counted as a single power of an $a_{k,q}$ coefficient, then the nonlinear constraints are homogeneous in powers of $a_{k,q}$.

Up to and including $\mathcal{O}(X^4)$, all $5$-point coefficients $w_{i,j}$ are fixed in terms of the 4-point Wilson coefficients $a_{k,q}$, but this is not the case at higher orders. At  $\mathcal{O}(X^5)$, 25 of the 26 independent 5-point Wilson coefficients are fixed in terms of the $a_{k,q}$'s while only a \textit{single} $w_{i,j}$ remains free. 
We can understand this freedom in the parameterization from  $\mathcal{O}(X^5)$ being the first order at which there exists a cyclically symmetric polynomial in the $X_{i,j}$ that vanishes at the locus of the splits \eqref{eq:5ptSplit};  therefore it drops out of all the constraints. At higher orders, the coefficient of any cyclically invariant polynomial  times this $\mathcal{O}(X^5)$ 
is likewise unfixed. 

\begin{table}[t]
\centering
\begin{tabular}{lc}
\multicolumn{1}{l|}{$k$}                  & Non-linear constraints on $a_{k,q}$        \\ \hline
\multicolumn{1}{l|}{\multirow{2}{*}{2}} & \multirow{2}{*}{$a_{2,1}= \frac{3}{2}a_{2,0} -\frac{1}{2g^2}a_{0,0}^2$} \\
\multicolumn{1}{l|}{}                   &                   \\ \hline
\multicolumn{1}{l|}{\multirow{2}{*}{3}} & \multirow{2}{*}{$a_{3,1}= 2 a_{3,0}-\frac{1}{g^2}a_{1,0} a_{0,0}.$} \\
\multicolumn{1}{l|}{}                   &                   \\ \hline
\multicolumn{1}{l|}{\multirow{2}{*}{4}} & \multirow{2}{*}{$a_{4,1}=
\frac{5}{2} a_{4,0} - 
\frac{1}{g^2} a_{0,0} a_{2,0}
-\frac{1}{2g^2}a_{1,0}^2\,, 
\quad 
a_{4,2}=
\frac{10}{3}a_{4,0}
-\frac{1}{g^2} a_{1,0}^2
-\frac{3}{2g^2} a_{2,0}a_{0,0}
+\frac{1}{6g^4}a_{0,0}^3$}    \\
\multicolumn{1}{l|}{}                   &                      \\ \hline
\multicolumn{1}{l|}{\multirow{4}{*}{5}} & \multirow{2}{*}{
$a_{5,1}
=
3 a_{5,0}
-\frac{1}{g^2} a_{2,0}a_{1,0}
-\frac{1}{g^2} a_{3,0}a_{0,0}\,, $}    \\
\multicolumn{1}{l|}{}                   &                      \\
\multicolumn{1}{l|}{}                   & \multirow{2}{*}{
$a_{5,2}=
5a_{5,0}
-\frac{5}{2g^2} a_{2,0}a_{1,0}
-\frac{2}{g^2} a_{3,0}a_{0,0}
+\frac{1}{2g^4}a_{1,0}a_{0,0}^2$}    \\
\multicolumn{1}{l|}{}                   &                      \\ \hline
\multicolumn{1}{l|}{\multirow{6}{*}{6}} & \multirow{2}{*}{$
a_{6,1}
=
\frac{7}{2}a_{6,0}
-\frac{1}{g^2} a_{3,0}a_{1,0}
-\frac{1}{g^2} a_{4,0}a_{0,0}
-\frac{1}{2g^2} a_{2,0}^2\,, $}    \\
\multicolumn{1}{l|}{}                   &                      \\
\multicolumn{1}{l|}{}                   & \multirow{2}{*}{$a_{6,2}
=7a_{6,0}
-\frac{3}{2g^2} a_{2,0}^2
-\frac{3}{g^2} a_{3,0}a_{1,0}
-\frac{5}{2g^2} a_{4,0}a_{0,0}
+\frac{1}{2g^4}a_{1,0}^2a_{0,0}
+\frac{1}{2g^4}a_{2,0}a_{0,0}^2$}    \\
\multicolumn{1}{l|}{}                   &                      \\
\multicolumn{1}{l|}{}                   & \multirow{2}{*}{$a_{6,3}
=
\frac{35}{4}a_{6,0}
-\frac{17}{8g^2} a_{2,0}^2
-\frac{4}{g^2} a_{3,0}a_{1,0}
-\frac{10}{3g^2} a_{4,0}a_{0,0}
+\frac{1}{g^4}a_{1,0}^2a_{0,0}
+\frac{3}{4g^4}a_{2,0}a_{0,0}^2
-\frac{1}{24g^6} a_{0,0}^4
$}    \\
\multicolumn{1}{l|}{}                   &                      \\             
\end{tabular}
\caption{\label{tab:Constraintsk6}Nonlinear constraints on the $4$-point Wilson coefficients, $a_{k,q}$, up to $k=6$, from imposing the 5-point splits \reef{eq:5ptSplitA}-\reef{eq:5ptSplitB}.}
\end{table}

A natural question is whether the splitting conditions at 6- and higher point further constrain the $a_{k,q}$'s and $w_{i,j}$'s.  
As discussed in Appendix \ref{app:LocalTerms}, the $6$-point splits (e.g.~$\mathcal{A}_6 \to \mathcal{A}_5 \times \mathcal{A}_4$) \textit{do not} impose any new constraints between the $a_{k,q}$'s, but they let us fix certain $5$-point coefficients. 

Before we proceed to the dispersion relations, let us  discuss standard examples of crossing symmetric scalar amplitudes in the context of nonlinear constraints in Table \ref{tab:Constraintsk6}.

\subsection{Example Amplitudes}\label{sec:examps}

In the Introduction, we discussed the beta function amplitude \reef{eq:4ptStringAmp} 
whose low-energy expansion has Wilson coefficients \reef{Venakqs} and  $g^2 = 1/\alpha'$.
It is easy to see that these $a_{k,q}$ satisfy the nonlinear constraints found from the 5-point splits.  For example, the $k=2$ constraint from Table \ref{tab:Constraintsk6} becomes (setting $\alpha'=1$ for simplicity)
\be
  \frac{\zeta_4}{4} = 
  \frac{3}{2} \zeta_4 
  - \frac{1}{2} 
  \big(\zeta_2 \big)^2
  ~~~~\text{i.e.}~~~~
  \frac{\pi^4}{4 \cdot 90} 
  = \frac{3}{2}\frac{\pi^4}{90}
  - \frac{1}{2} \frac{\pi^4}{36} 
  ~~~~
  \checkmark
\ee
Similarly, one can explicitly check that the higher-$k$ nonlinear constraints in Table \ref{tab:Constraintsk6} are solved by the beta-function Wilson coefficients.

Two amplitudes that are often at the boundaries of bootstrap regions are the Infinite Spin Tower (IST)
and the massive scalar exchange amplitudes:  
\be\label{eq:istamp}
  \begin{split}
   &
  \tilde{\mathcal{A}}_4^\text{IST} = -\frac{g^2}{su}+
   \frac{\lambda^2}{(M^2 - s)(M^2 - u)}\,,
   \\
   &
   a_{0,0}= \frac{\lambda^2}{M^4}\,,
   ~~
   a_{1,0}= \frac{\lambda^2}{M^6}\,,
   ~~
   a_{2,0}= a_{2,1}=\frac{\lambda^2}{M^8} \,,
   ~~
   a_{3,0}= a_{3,1}=\frac{\lambda^2}{M^{10}} \,,
   \ldots
   \end{split}
\ee
and
\be\label{eq:SVamp}
  \begin{split}
   &
   \tilde{\mathcal{A}}_4^\text{scalar} = -\frac{g^2}{su}+
   \frac{\lambda^2}{M^2 - s}
   +\frac{\lambda^2}{M^2 - u} \,,
   \\
   &
   a_{0,0}= \frac{2\lambda^2}{M^4}\,,
   ~~
   a_{1,0}= \frac{\lambda^2}{M^6}\,,
   ~~
   a_{2,0}= \frac{\lambda^2}{M^8} \,,
   ~~
   a_{2,1}=0\,,
   ~~
   a_{3,0}= \frac{\lambda^2}{M^{10}} \,,
   ~~
   a_{3,1}=0\,,
   \ldots
   \end{split}
\ee
Upon multiplication by an overall factor of $-t$, as in \reef{AtildeDef}, the amplitudes $\mathcal{A}_4$ satisfy the hidden zero condition, $\mathcal{A}_4 \to 0$ for $t \to 0$. 
In addition, the $a_{k,q}$ of the Hidden Zero IST amplitude, $\mathcal{A}^\text{HZIST}= -t \tilde{\mathcal{A}}_4^\text{IST}$, solves the nonlinear relations in Table \ref{tab:Constraintsk6} (and those at higher $k$ too) when we choose $\lambda = g$, so that means that this amplitude {\em is compatible} with the 5-point splitting constraints.\footnote{The fact that this IST amplitude satisfies these constraints makes it plausible that there could be an $n$-point extension of for such an amplitude with infinitely many spins at the same mass that satisfies the splitting constraints. Adding the requirement that the amplitude have these splits could be an interesting criterion for trying to write down a unique expression for such an $n$-point amplitude.} 

In contrast, the scalar-vector exchange amplitude 
$\mathcal{A}^\text{SV}= -t \tilde{\mathcal{A}}_4^\text{scalar}$
is {\em incompatible} with the 5-point splitting constraints. Specifically, the  $k=2$ nonlinear constraint require $\lambda^2 = 3g^2/4$ whereas the $k=3$ requirement is solved for $\lambda^2 = g^2$. Hence, no nonzero choice of 
coupling $\lambda$ between the massive states exchanged and the external massless states makes the amplitude compatible with the 5-point splitting formulas. 
Thus, upon implementing the nonlinear constraints in the numerical bootstrap, we should find that the $\mathcal{A}^\text{SV}$ is excluded. This is indeed the case, as we show in the following section.

\section{The Bootstrap}
\label{sec:Bootstrap}
In this section, we give the essential information for understanding how the splits reduce the allowed space of Wilson coefficients: dispersion relations, standard null constraints, and a recipe for incorporating the non-linear relations from the splitting conditions as linearized constraints. We present the simplest bounds in order to display the significant effects of imposing constraints from higher-point amplitudes on the 4-point Wilson coefficients. 

The bootstrap bounds depend on the positivity of the imaginary part of partial wave coefficients $\rho_j(s)$ as given in  \eqref{eq:specden} implied by unitarity. However, the assumption that $\mathcal{A}_4(s,u)= (s+u) \tilde{\mathcal{A}}_4(s,u)$ has a positive partial wave expansion does not imply a similar statement for $\tilde{\mathcal{A}}_4$. Hence we cannot strip off the $(s+u)$-factor and bootstrap $\tilde{\mathcal{A}}_4$ without making the additional assumption that $\tilde{\mathcal{A}}_4$ has a positive partial waves.
Instead we bootstrap $\mathcal{A}_4(s,u)$ directly, assuming the improved Regge behavior \eqref{eq:bounduandtfixed}.

We are going to determine  bounds on ratios of the 4-point  coefficients $a_{k,q}$. However, it is technically simpler to set up the bootstrap problem in terms of a different set of coefficients $c_{k,q}$ that have simple dispersive representations and are defined by the low-energy expansion 
\begin{equation}
  \label{A4wckq}
    \mathcal{A}_4^{\text{EFT}}(s,u) = -\frac{g^2}{s} -\frac{g^2}{u} + \mathcal{A}_4^{\text{C}} , \quad \text{with} \quad  \mathcal{A}_4^{\text{C}} =\sum_{k=0}^{\infty}\sum_{q=0}^{k} c_{k,q} s^{k-q} u^{q},
\end{equation}
These coefficients have to satisfy crossing, $c_{k,k-q} = c_{k,q}$, and they must also obey linear constraints 
\begin{equation}
    c_{0,0} = 0, \quad \quad \frac{c_{2,1}}{2} = c_{2,0}, \quad  \quad\frac{c_{4,2}}{2} = c_{4,1} - c_{4,0}, \quad \quad \cdots 
    \label{eq:constraintsLowk}
\end{equation}
in order for the amplitude to have the hidden zero $\mathcal{A}_4^{\text{EFT}}(s,-s) = 0$.

The $c_{k,q}$ are related to the $a_{k,q}$ from \reef{eq:WCoeffZ} by a simple linear transformation described in Section \ref{s:asandcs}.  
By matching these two expansions, we are also able to derive the general form of the constraints on the $c_{k,q}$ coefficients by the hidden zero (extending \eqref{eq:constraintsLowk} to higher $k$). 
In Section \ref{s:Disp_Cs} we discuss the dispersion relations for the $c_{k,q}$ coefficients and in Section \ref{s:NonLin} explain how to implement the nonlinear split relations as constraints on the spectral density $\rho_j$.

\subsection{From $c_{k,q}$'s to $a_{k,q}$'s}
\label{s:asandcs}

The two low-energy expansions, \reef{eq:WCoeffZ} and \reef{A4wckq} describe the same amplitude and equating them order-by-order in the Mandelstam expansion, we extract the relationship between the $a_{k,q}$ and $c_{k,q}$:
\begin{equation}
\begin{aligned}
    &c_{0,0} =0, \\
    &c_{k,0} = a_{k-1,0}, 
\end{aligned} \quad \quad 
\begin{aligned}
    &c_{k,k/2} = 2 a_{k-1,k/2-1}, \\
    &c_{k,q} = a_{k-1,q-1} + a_{k-1,q}, \quad (q\neq \{0,k/2\}).
\end{aligned} 
\label{eq:ctoa}
\end{equation}
We can invert the last equation to write the $a_{k,q}$ coefficients in terms of the $c_{k,q}$ coefficients:
\begin{equation}
    a_{k,q} = (-1)^q \sum_{q^\prime =0}^{q} (-1)^{q^\prime} c_{k+1,q^\prime}, \quad \text{for }q\neq 0.
    \label{eq:atoc}
\end{equation}
This imposes a constraint between the $c_{k,q}$'s which comes from setting $k = k^\prime-1$ and $q = k^\prime/2-1$ (with $k^\prime$ even) in the equation above, and equating it with the first line on the right of \eqref{eq:ctoa}:
\begin{equation}
\frac{c_{k^\prime,k^\prime/2}}{2} = (-1)^{k^\prime/2-1} \sum_{q^\prime =0}^{k^\prime/2-1} (-1)^{q^\prime} c_{k^\prime,q^\prime}\,.
    \label{eq:constraintsc}
\end{equation}
This is nothing but the hidden zero conditions; one can easily check that \reef{eq:constraintsc} precisely gives the constraints in \eqref{eq:constraintsLowk} for $k^\prime =0,2,4$.
Therefore, we have that the EFT for the $4$-point amplitude with the hidden zero can be parametrized in terms of the $c_{k,q}$ for $q<k/2$.

\subsection{Dispersion Relations and Null Constraints}
\label{s:Disp_Cs}

Following a standard contour deformation argument, detailed for example in \cite{Berman:2023fes}, we derive dispersion relations for the Wilson coefficients $c_{k,q}$ of the 4-point EFT amplitude \reef{A4wckq}. To briefly recap, each $c_{k,q}$ is picked out by taking derivatives of a simple contour integral in the complex-$s$ plane with fixed $u < 0$:
\begin{align}\label{formaldefinition}
    c_{k,q} = \bigg(\frac{1}{q!}\frac{\partial^q}{\partial u^q}\int_{\mathcal{C}_0} \frac{ds}{2\pi i}\frac{\mathcal{A}_4^{\text{EFT}}(s,u)}{s^{k-q+1}}\bigg)\bigg|_{u= 0}\, .
\end{align}
Here $\mathcal{C}_0$ is a contour around $s = 0$ that does not extend above $s = M_{\gap}^2$. By deforming the contour to wrap the $s$-axis with $s \ge M_{\gap}^2$ on the positive real $s$-axis, a simple dispersive expression for each coefficient is derived.
By the assumption of the improved Regge behavior \reef{eq:bounduandtfixed} at fixed $u<0$, the contribution from the contour at infinity vanishes for all $k$ and $q$ with $k-q \ge 1$ and hence we have the {\em once-subtracted dispersion relations} 
\begin{equation}
\label{finalequ}
c_{k,q}\,=\,\sum_{j=0}^{\infty}\int_{1}^\infty dy \,f_{j}(y) \,y^{-k} v_{j,q}  
~=~ 
\Big\langle y^{-k} v_{j,q}  \Big\rangle_1 \,
\ 
~~~~
\text{for}~~k-q \ge 1\,,
\end{equation}
Here $y = s/M_{\gap}^2$, $f_{j}(y) =  y^{-d/2} \rho_{j}(M_{\gap}^2y) \ge 0$ and $v_{j,q}$ is the $q$-th derivative of the Gegenbauer polynomials,
\begin{equation}
    v_{j,q} = \frac{1}{q!} \frac{\partial^q}{\partial x^q} \mathcal{P}_j^{(d)}(1+2x) \bigg \vert_{x=0}\,.
\end{equation} 
The bracket notation, $\Big\langle \cdot \Big\rangle_{\mu_c}$ is shorthand for the dispersive integral
\be 
 \label{defbracket}
\Big\langle Q\Big\rangle_{\mu_c}
= \sum_{j=0}^{\infty}\int_{\mu_c}^\infty dy \,f_{j}(y) \,Q(j,y) \,.
\ee
Here $\mu_c$ denotes a cutoff scale that signifies that we are agnostic about the spectrum above $\mu_c M_\gap^2$. 
Specifically, without spectrum assumptions (beyond the existence of the mass gap) we have $\mu_c =1$, as in \reef{finalequ}.

\subsection*{Null Constraints} 
There are two sets of {\em null constraints} on the spectral density $f_j(y)$ in \reef{finalequ}. The first comes from the simple $s\leftrightarrow u$ crossing symmetry of the amplitude
\begin{align}
  \label{basicCrossing}
    0 = c_{k,q} - c_{k,k-q} \, .
\end{align}
The second set of null constraints \cite{Albert:2022oes} arises from relating the constant $u<0$ dispersion relations to those obtained with constant $t<0$. 
Writing the low-energy expansion of the amplitude in terms of $s$ and $t$, we have
\begin{align}
    \mathcal{A}_4^{\text{EFT}}(s,-s-t) = \sum_{k=0}^{\infty}\sum_{q=0}^{k}b_{k,q}s^{k-q}t^{q} \,,
~~~~
\text{with}
~~~~
    b_{k,q} = \bigg(\frac{1}{q!}\frac{\partial^{q}}{\partial t^{q}}\int_{\mathcal{C}_0} \frac{ds}{2\pi i}\frac{\mathcal{A}_4^{\text{EFT}}(s,-s-t)}{s^{k-q+1}}\bigg)\bigg|_{t= 0}\, .
\end{align}
Deforming the contour as before, the contribution from infinity vanishes for $k-q \ge 1$ (due to the second Regge limit in equation \reef{eq:bounduandtfixed}), and we find
\be
b_{k,q} = \sum_{j=0}^\infty
\int_1^\infty dy\,
f_{j}(y) \,y^{-k} (-1)^j
\bigg[
  v_{j,q} +
  (-1)^k
  \sum_{q'=0}^{q}
  (-1)^{q'} {{k-q'} \choose {q-q'} }v_{j,q'}
\bigg] \,.
\ee 
The $c_{k,q}$ and $b_{k,q}$ coefficients are directly related by 
\be  
 \label{asfrombs}
  c_{k,q}= 
  \sum_{q'=q}^k (-1)^{q'} 
  {{q'} \choose {q}} b_{k,q'} \,,
\ee 
and this gives a set of null relations independent from \reef{basicCrossing} for $0 \le q \le \left\lceil k/2 \right\rceil$. 
However, equation \reef{asfrombs} involves $b_{k,k}$, which we do not have access to via the once-subtracted dispersion relations \reef{finalequ}. 
Hence, for each $k$, we use the $c_{k,0}$ relation to eliminate the $b_{k,k}$ coefficient from the remaining conditions in \reef{asfrombs}, and the result are the null constraints
\be\label{asfrombs1sdr}
   0 = c_{k,q} - {k \choose q} c_{k,0}
   - \sum_{q'=q}^{k-1} (-1)^{q'} {{q'} \choose {q}} b_{k,q'}
   + {k \choose q} 
   \sum_{q'=0}^{k-1} (-1)^{q'}
   b_{k,q'} \, .
\ee
for $1 \le q \le \left\lceil k/2 \right\rceil$. 

Let us now proceed to briefly outline the numerical implementation of the null constraints.

\subsection*{Numerical Bounds: Implementation}
\label{s:NumImpl}

Multiplying the dispersive representation \reef{finalequ} by a positive number does not change positivity of the spectral density, so the bounding problem is projective and we can only bound ratios of  Wilson coefficients.  It is natural to bound the $a_{k,q}$'s relative to the lowest accessible coefficient, e.g.~
\be
  \frac{a_{k,q}}{a_{0,0}}\,.
\ee
This translates to a linear combination of $c_{k,q}$'s divided by $c_{1,0}$.

\begin{figure}[t]
\centering
\includegraphics[width=0.6\textwidth]{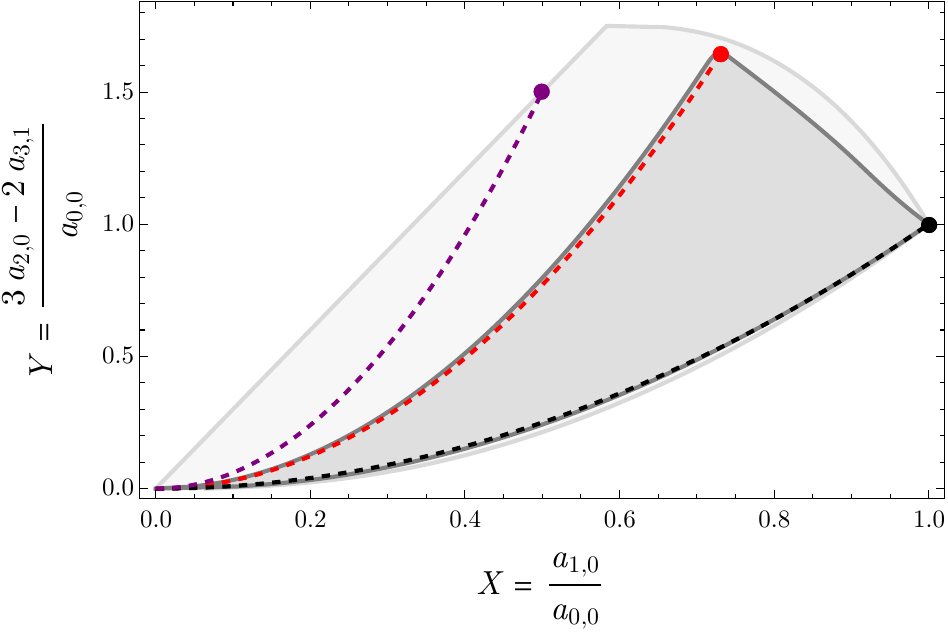}
\caption{Convex hidden zero region vs.~non-convex region from nonlinear splitting conditions. The $d = 10$ general bounds in the $(X,Y) = \big(a_{1,0}/a_{0,0},(3a_{2,0}-2a_{2,1})/a_{0,0}\big)$ plane at $k_{\max} = 12$ without (light gray) and with (gray) the nonlinear splitting conditions \reef{nonlinAk8Yonly} imposed. The purple dashed line is the massive scalar-vector amplitude \reef{eq:SVamp} with $M_{\gap}^2/M^2 \leq 1$, the red dashed line is the beta function amplitude \reef{eq:4ptStringAmp} with $\alpha'M_{\text{gap}}^2 \leq 1$, and the black dashed line is the spin tower amplitude \reef{eq:istamp} with $M_{\gap}^2/M^2 \leq 1$.}
\label{fig:genbds}
\end{figure}

It follows directly from the dispersion relations \reef{finalequ} that $0 \le c_{k,q}/c_{1,0} \le 1$. Moreover, using the hidden zero condition, we find that the same holds for the ratio $a_{k,q}/a_{0,0}$. To obtain optimal two-sided bounds on ratios of Wilson coefficients, the $a_{k,q}$'s are translated to the $c_{k,q}$ via the linear map \reef{eq:atoc} and the dispersive expressions \reef{finalequ} for the $c_{k,q}$ are used to define a semidefinite optimization problem that we solve numerically using SDPB \cite{Simmons-Duffin:2015qma}.  As an example, 
scanning over the allowed values of $0 \le X \equiv a_{1,0}/a_{0,0} =c_{2,0}/c_{1,0}\le 1$, we can minimize and maximize any linear combination of other ratios $a_{k,q}/a_{0,0}$ to obtain the allowed region in the  $(X,Y)$-plane. 
Without additional assumptions, this always gives a convex region; an example is the light gray region in Figure \ref{fig:genbds}. 

We impose constraints with $k \le k_\text{max}$, where the value of $k_\text{max}$ reflects the derivative order in the low-energy effective expansion. 
Specifically, we denote $k_{\max}$ in terms of the $a_{k,q}$ counting of $k$,
but the numerics are performed via the $c_{k,q}$'s with $k$ offset by one due to the $(s+u)$ factor in the amplitude \reef{eq:WCoeffZ} vs.~the expansion in \reef{A4wckq}.
For example $k_{\max} = 4$ means that we impose the null constraints \eqref{eq:constraintsLowk}, \eqref{eq:constraintsc}, and \eqref{asfrombs1sdr} up to $k = 5$, but when we impose the nonlinear constraints in Table \ref{tab:Constraintsk6} it is up to $k=4$. As $k_\text{max}$ is increased, the bounds become strictly stronger.

\subsection{Imposing Nonlinear Constraints}
\label{s:NonLin}

A crucial feature of the constraints from the $5$-point splits from Section \ref{s:HZEFT} is the fact that they are nonlinear in the 4-point Wilson coefficients $a_{k,q}$.
These nonlinear relations can be treated as constraints on the spectral density $f_{j}(y)$ via the linear relation \reef{eq:ctoa} between the $a_{k,q}$ to the $c_{k,q}$. However, two technical issues  present independent challenges for implementing the nonlinear constraints from Table \ref{tab:Constraintsk6} directly. The first, which is overcome easily, is that the 3-point coupling $g$ of the massless state appears in the constraints. This cubic coupling cannot be written dispersively with our assumed Regge behavior. 
Instead, we use the $k = 2$ nonlinear constraint to solve for $g^2$, obtaining 
\be
  \label{gSQ}
    g^2 = \frac{a_{0,0}^2}{3a_{2,0}-2a_{2,1}}\, ,
\ee
and use this to eliminate $g^2$ in the higher $k$ non-linear relations to get constraints purely in terms of the Wilson coefficients. As a result, the nonlinear relations become homogeneous in scalings of the $a_{k,q}$. In particular, for $k=3$  we find 
\be
  \label{k3nonlinA}
   \begin{array}{llrcl}
   k=3\!: &~~~~& 
   a_{1, 0} \big(3 a_{2, 0} - 2 a_{2, 1}\big) 
   &=& 
   a_{0, 0} \big(2 a_{3, 0} - a_{3, 1}\big) \,.
   \end{array}
\ee
The second issue is that the nonlinearity means the bootstrap with splitting conditions cannot be phrased directly as a semidefinite optimization problem: the conditions do not correspond to the semidefinite positivity of matrices. In fact, the splitting conditions are not even convex. In order to implement these splitting conditions in general, then, we would need to use some generic, non-convex optimization scheme.\footnote{For example, one could bound the space with the splitting conditions imposed directly on the Wilson coefficients using the analytic bounds described in \cite{Arkani-Hamed:2020blm,Chiang:2021ziz}. However, this method quickly becomes computationally inefficient and it is also not known how to implement the fixed-$t$ null constraints purely in terms of $a_{k,q}$ coefficients.} 

In this paper, we resolve this issue in a more straightforward way: we effectively linearize (a subset of) the nonlinear constraints by scanning over certain combinations of coefficients. Specifically, introducing 
\begin{equation}
\label{defXY}
    X \equiv \frac{a_{1,0}}{a_{0,0}} = \frac{c_{2,0}}{c_{1,0}} 
    \,, ~~~~~~~
    Y \equiv \frac{3a_{2,0}-2a_{2,1}}{a_{0,0}} = \frac{5c_{3,0}-2c_{3,1}}{c_{1,0}}
     \, ,
\end{equation}
we can write the condition \reef{k3nonlinA} in three different ways:
\begin{eqnarray}
  \label{k3nonlinB1}
   k=3\!: &~~~~&  
   X \,
   \big( 3a_{2,0}-2a_{2,1}\big)  
   = 
   2a_{3, 0} - a_{3, 1} \,,
   \\
     \label{k3nonlinB2}
   ~~~\text{or}&~~~~&  
   a_{1,0} \,Y
   = 
   2a_{3, 0} - a_{3, 1} \,,
   \\
     \label{k3nonlinB3}
   ~~~\text{or}&~~~~& 
   X \,Y a_{0,0}
   = 
   2a_{3, 0} - a_{3, 1} \,.
\end{eqnarray}
Each of these equivalent representations of the $k=3$ constraint becomes a linear constraint when we input explicit values of either $X$, or $Y$, or both, while we minimize/maximize another variable $a_{k,q}/a_{0,0}$ (or a linear combination of such ratios). Thus scanning over values of $X$ and/or $Y$, 
as outlined at the end of Section \ref{s:NumImpl}, allows us to implement the $k=3$ constraint in one of the above forms. 

At $k=4$, we write the nonlinear constraints in terms of $X$ and $Y$ as
\be
  \label{k4nonlinA}
   \begin{array}{llrcl}
   k=4\!: &~~~~& 
   6 X^2 Y a_{0,0} - Y^2 a_{0,0} + 9 Y a_{2,0} - 20 a_{4,0} + 
 6 a_{4,2}&=& 0
   \,, \\
    &~~~~& 
   X^2 Y a_{0,0} + 2 Y a_{2,0} 
   - 5 a_{4, 0} + 2 a_{4,1} &=& 0
 \,.
   \end{array}
\ee
These two constraints are both linearized if we scan over both $X$ and $Y$; however, scanning over just $Y$, we can only impose the linear combination that eliminates the $X^2Y$-term, namely
\be
 \label{k4nonlinB}
- Y^2 a_{0,0} - 3 Y a_{2,0} 
+ 10 a_{4,0} 
- 12 a_{4,1}
+ 6 a_{4,2} 
  = 0\,.
\ee
Likewise, we find one linear combination of the splitting constraints can be linearized for given $Y$ at $k=5,6$ and $8$, however, this does not seem possible at $k=7,9,$ and $10$.  Appendix \ref{app:nonlin} gives further details and the full set of $Y$-scan constraints for $k \le 8$ are summarized in \reef{nonlinAk8Yonly}. 
It is simpler to compute bounds when scanning only over a single variable, so we proceed with that in this section, while bounds resulting from scans over both $X$ and $Y$ are discussed in Section \ref{sec:spectrum}.

\subsection{Non-Convex Allowed Regions}
\label{s:basicbounds}

Consider the 2d plane $(X,Y)$, with $X$ and $Y$ defined in \reef{defXY}. Including the standard null constraints \reef{basicCrossing} and \reef{asfrombs1sdr} up to $k_\text{max}=12$, we use SDPB to determine the bounds on $X$ and $Y$ numerically and find that they have to be in the lightgray region shown in Figure \ref{fig:genbds}.
It is clearly convex, as guaranteed by the linearity of the positivity conditions. 

The moment we include even just one nonlinear constraint, e.g.~the $k=3$ constraint in \reef{k3nonlinB2}, convexity is lost. Scanning over $Y$, we impose the constraints in \reef{nonlinAk8Yonly} for  $k$ up to $k_\text{max}$ and compute 
the minimum and maximum allowed values of $X$. 
The resulting allowed region is shown as the non-convex darkgray region in Figure \ref{fig:genbds}; it was computed at $k_\text{max}=12$ with all five nonlinear constraints imposed \reef{nonlinAk8Yonly}. Its non-convexity is in sharp contrast to the lightgray convex region. 
To reiterate, the difference is that the convex lightgray region is computed assuming that the 4-point amplitude has a hidden zero, but without imposing the 5-point splitting conditions. With the 5-point splitting conditions, only values of $X$t and $Y$ in the non-convex darkgray region are permissible.

In Section \ref{sec:examps}, we discuss three example amplitudes. First, the hidden zero scalar-vector exchange amplitude \reef{eq:SVamp} is shown in blue in Figure \ref{fig:genbds} and it is  clearly excluded from the dark gray region. This was expected because the scalar-vector amplitude does not satisfy the nonlinear constraints.

Second, the infinite spin tower amplitude \eqref{eq:istamp} with mass $M$ has 
\be (X,Y) = (M_\text{gap}^2/M^2,M_\text{gap}^4/M^4)\ee 
and does satisfy the nonlinear splitting constraints. Varying $M \ge  M_\text{gap}$, this traces out the black dashed curve in Figure \ref{fig:genbds}. It lies within the dark gray region and we anticipate that the lower bootstrap bound  saturates to 
this curve in the limit $k_\text{max} \to \infty$. 

Third, Section \ref{sec:examps} discussed the beta function amplitude (\ref{eq:4ptStringAmp}). It has 
\be
(X,Y) = (\alpha'\zeta_3/\zeta_2,\alpha'^2\zeta_2)
\ee
and is located at the black dot for $\alpha'M_{\gap}^2 =1$ while varying $\alpha'M_{\gap}^2 \ge 1$ gives the red dashed curve. The darkgray allowed region has a sharp corner close to the string with $\alpha'M_{\gap}^2 =1$, i.e.~when the first massive state of the string with mass $1/\alpha'$ is located at the mass gap. We expect that the upper bound on the dark gray region saturates to the red string curve when $k_\text{max}$ increases. 

Finally, the dark gray region has a bound between the string with $\alpha'M_{\gap}^2 =1$ (black dot) and the infinite spin tower at the mass gap (red dot). This bound is practically a straight line so one might naively think that the extremal amplitudes along this bound are positive linear combinations of these two extremal amplitudes. This is not the case.  
Unlike in the standard positivity bootstrap problems, sums of consistent amplitudes are \textit{not} allowed when we impose the nonlinear splitting conditions. For example, while the string amplitude with any choice of $\alpha'M_{\text{gap}}^2 \leq 1$ is valid on its own, sums of two or more such amplitudes with different choices of $\alpha'M_{\text{gap}}^2$ do not live within the allowed region. (Their Wilson coefficients would be in the convex hull of the red curve in Figure \ref{fig:genbds}, but that is clearly ruled out by the non-convexity of the darkgray region.)  As a consequence, the existence of any allowed amplitudes in the region between the red and the black curves in Figure \ref{fig:genbds} 
is highly non-trivial, but we have tested that the bootstrap does not rule out points in the interior of the darkgray region. Since the amplitudes in the interior of the darkgray region or on its straight edge cannot come from simple sums of the two analytic amplitudes \reef{eq:4ptStringAmp} and \reef{eq:istamp}, they must themselves correspond to unique amplitudes. In particular, the top right boundary of the region must come from amplitudes with particle content precisely at the mass gap. We do not have analytic expressions for these  amplitudes, but, as we describe in the next section, they  appear to share a common feature: an infinite set of spinning states exactly at the mass gap $M_\text{gap}$. The beta function amplitude is the only one without this feature, making it unique among the set of amplitudes satisfying the splitting conditions!

\section{Splitting Bootstrap with Spectrum Input}
\label{sec:spectrum}
In this section, we present evidence that there are only two kinds of amplitudes that actually satisfy the splitting conditions: those with infinite towers of spins at finite mass and the string beta function amplitude. We show that excluding infinite spin towers at the mass-gap, isolates the string amplitude within a tiny  shrinking island.

\subsection{Bootstrap with Spectrum Input}
\label{s:spectruminput}
Separating amplitudes with and without spin towers can be done in different ways by making basic assumptions about the spectral density $f_{j}(y)$ in the dispersion relations \reef{finalequ}. One option is to impose the existence of a linear leading Regge trajectory with finite slope, as done for example in \cite{Haring:2023zwu}. 
Here, we are not necessarily concerned with directly eliminating the possibility of spin towers, since it appears that can appear as a solution to our bootstrap conditions. Instead, we make the less restrictive assumption that there is not an infinite tower of spinning states exchanged exactly at the mass gap $M_\gap$.

To do so, we introduce a cutoff scale $\mu_c M_{\gap}^2$ such that the spectral density $f_{j}(y)$ in \eqref{finalequ} has no support on the open interval $1<y< \mu_c$; this means that there are no states in the $s$-channel with mass-squared between $M_{\gap}^2$ and $\mu_c M_{\gap}^2$. 
We can then rewrite \eqref{finalequ} in two pieces, separating the contributions from $y = 1$ and from $y \geq \mu_c$ \cite{Albert:2022oes,Albert:2023seb,Berman:2024wyt}:
\be
\label{ckqdispWstates1}
    c_{k,q} = \sum_{j=0}^{j_\text{max}}|\lambda_{j}|^2v_{j,q} + \<y^{-k}v_{j,q}\>_{\mu_c}\, .
\ee
Here we have made the key  assumption that only a finite number of spin states appear at the mass gap. 
The $\lambda_{j}$ are proportional to the 3-point coupling between the two massless external scalars and the exchanged spin-$j$ state with mass $M_{\gap}^2$.

For weakly-coupled amplitudes with once-subtracted dispersion relations, such as $\mathcal{A}_4$, it has been proven analytically in \cite{Berman:2024kdh,Berman:2024owc} that the maximally allowed spin at the mass gap is spin 1, i.e.~we must have $j_\text{max}=1$ in \reef{ckqdispWstates1}. 
Thus, without loss of generality, the assumption that there is only a finite number of spin states exchanged at the mass gap is equivalent to assuming that there are only scalar and vector exchanges at $y=1$, and hence we have 
\begin{align}
    c_{k,q} = |\lambda_{0}|^2 v_{0,q} + |\lambda_{1}|^2 v_{1,q}+ \<y^{-k}v_{j,q}\>_{\mu_c}\, .
\end{align}
Further, the hidden zero constraint requires that
\begin{align}
    A(s,-s) = 0\,
\end{align}
so, setting $s+u=0$ in the partial wave expansion \reef{eq:pwexp}, we find
\begin{align}
\label{ImAHZ}
0 = \text{Im}[A(s,-s)] = \sum_{j=0}n_j^{(d)}\text{Im}[a_j(s)]\mathcal{P}_j^{(d)}(-1) = \pi s^{(d-4)/2}\sum_j(-1)^j\rho_j(s)\, .
\end{align}
The contribution of the tree-level exchange of a spin-$j$ state at $s = m^2$ is
\be
\rho_j(s)\supset
    \delta(s-m^2) \,
    |\lambda_{n,j}|^{2}\,,
    \,
\ee
so that upon integrating \reef{ImAHZ} from $(M_{\gap}^2-\epsilon,M_\gap^2+\epsilon)$ we find that these couplings to the states at the gap must satisfy
\begin{align}
    0 = \pi ~M_{\gap}^{d-4}(|\lambda_0|^2-|\lambda_1|^2)\,.
\end{align}
We conclude that we must have $|\lambda_0|^2=|\lambda_1|^2$ for 4-point amplitudes with the hidden zero. Of course, 
this does not work if there 
are an infinite set of states at the mass gap, i.e.~if we allow $j_\text{max}=\infty$, 
so it need not apply for the IST amplitude (and indeed the IST has $|\lambda_0|^2\ne |\lambda_1|^2$). In the following, we impose the condition $|\lambda_0|^2=|\lambda_1|^2$ numerically by enforcing that the contributions from the scalar and spin-one states at the mass gap to the dispersion relations are equal.

Above $\mu_c$, any unitary spectrum is allowed, including those with spin towers. The space of theories with no contribution from the scalar/vector at the mass gap (i.e.~$\lambda_{0}=0, \lambda_{1}=0$) is easily identified by simply scaling down the full parameter space by powers of $\mu_c>1$ \cite{Berman:2024wyt,Berman:2024eid}. The remaining allowed space corresponds to only theories with nonzero contribution from the scalars at the gap. 

In the following two subsections, we study the resulting allowed regions in two different cases: the string-inspired case where the spacetime dimension is $d=10$ and the cutoff $\mu_c = 2$ as well as an example of the more generic case of $d=4$ and 
$\mu_c = 1.2$.

\subsection{Bifurcating Regions: $d=10$ and Cutoff 2}
\label{s:D10islands}

The beta function amplitude \reef{eq:4ptStringAmp} encodes exchange of an infinite set of states uniformly spaced in mass with states at $M^2_n = n/\alpha'$ for $n=0,1,2,\ldots$ and organized in Regge trajectories with slope 1. At the lowest mass, $M^2_1 = 1/\alpha'$, there is a scalar and a vector. 
The critical dimension of the beta function amplitude is $d=10$ \cite{Chiang:2023bst,Berman:2024wyt,Albert:2024yap}; the amplitude is unitary for $3 \le d\le 10$ but non-unitary for $d>10$. This means that in $d=10$, the beta function amplitude is borderline violating unitary and that gives the best opportunity for bootstrapping it. Therefore we set $d=10$ in this section.

Let us now assume the absence of an infinite spin tower at the mass gap. As discussed in Section \ref{s:spectruminput},  
 with a cutoff $\mu_c>1$ the only states allowed at the mass gap are a scalar and a vector, so assuming these are the only states at $M_\text{gap}$ does not represent any new assumptions beyond not allowing an infinite number of states there. Since the beta function amplitude has its second massive state at $2/\alpha$, we make the string inspired assumption of setting the cutoff to $\mu_c=2$, thereby excluding any states in the open interval between $M_\text{gap}^2$ and $2M_\text{gap}^2$. To summarize,
we make the assumption
\be
 \label{spectrum10}
 d=10\,,~~\text{finite \#{ of states} at $M_\text{gap}^2$,~~cutoff $\mu_c=2$.}
\ee
Let us start by simply imposing the $k=3$ nonlinear splitting constraints given in \reef{k3nonlinB2} for the choice \reef{spectrum10}. 
Remarkably, the allowed region then \textit{bifurcates} into two separate islands!

\begin{figure}[t]
    \centering
    \includegraphics[width=\linewidth]{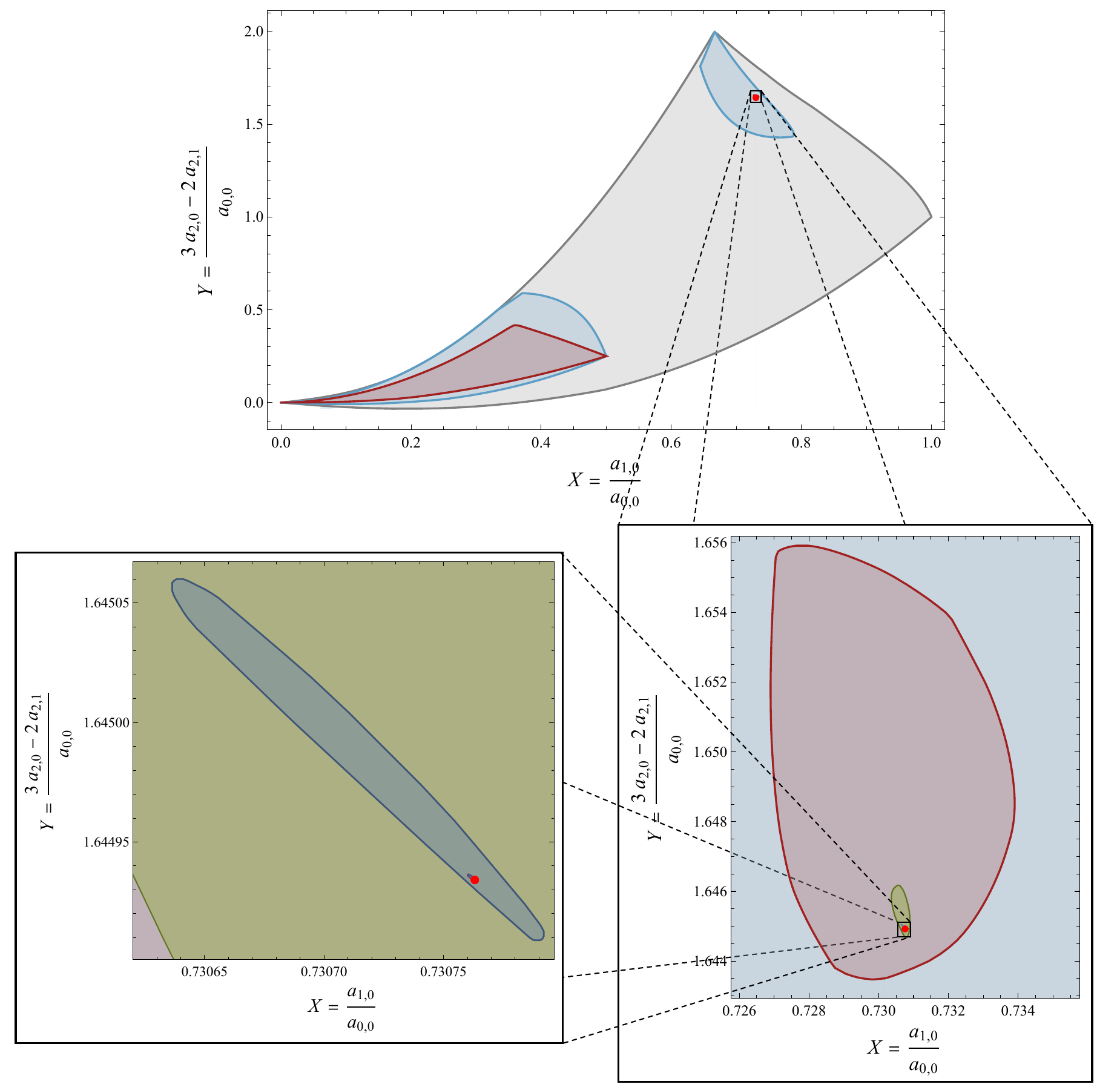}
    \caption{Allowed regions in the $(X,Y)$-plane for $d=10$. 
    {\bf Top:} Using only a single nonlinear constraint \reef{k3nonlinB2} at $k_{\max}=3$, the {\em gray region} are the bounds with no spectrum input  while  {\em light blue} shows the bifurcation of the region when we assume a finite number of states at the mass gap and cutoff $\mu_c=2$.  The {\em maroon region} is computed with $k_{\max}=6$, $\mu_c = 2$ and the three nonlinear constraints with $k\le 6$ given in \eqref{nonlinAk8Yonly}, performing a single-scan over $Y$. The red dot is the string amplitude (with $\alpha^\prime M^2_{\text{gap}} =1$). 
    {\bf Bottom Right:} Zoom-in on the  {\em maroon} single-scan $k_{\max}=6$. The {\em green island} corresponds to $k_{\max}=6$ island obtained from a double-scan over $X$ and $Y$ with seven nonlinear constraints (given in \eqref{nonlinA}-\eqref{nonlinAk6XY}) with $k \le 6$ imposed. 
    {\bf Bottom Left:} Further zoom on the green island shows the $k_{\max}=8$ {\em dark blue island} from double-scan over $X$ and $Y$ with eleven nonlinear constraints and $k \le 8$. (See Appendix \ref{app:nonlin} for details of the double-scan.) Barely visible but slightly extending out behind the red dot, we also show the tiny $k_{\max} = 10$ island in purple.}
    \label{fig:Zoom10d}
\end{figure}

We show the bifurcation in the top plot in Figure \ref{fig:Zoom10d}. For reference, we show in gray the $d=10$ non-convex region resulting from imposing the nonlinear splitting constraint \reef{k3nonlinB2} for just $k=3$ but without the spectrum input \reef{spectrum10}. (This is similar to the dark gray region in Figure \ref{fig:genbds}, but now only with $k_{\max}=3$). 
Once we require the spectrum input \reef{spectrum10} along with the $k=3$ nonlinear constraint, we get the {\em two} separated blue regions (obtained  at $k_{\max}=3$). The bounds on these regions are determined by scanning over the range of $Y$ while optimizing $X$ and finding that SDPB only converges to a solution in the two separate regions. The upper island includes the beta function amplitude indicated by the red dot. 

Increasing the number of constraints to $k_{\max} = 6$ (imposing  $Y$-scan constraints from \reef{nonlinAk8Yonly} up to $k_{\max} = 6$), the allowed parameter space shrinks to the maroon regions shown in Figure  \ref{fig:Zoom10d}. The island around the beta function amplitude is not visible in the large-scale plot, but is shown the zoomed-in plot on the bottom right in Figure \ref{fig:Zoom10d}.
The lower maroon region can be  understood as the scaled down version of the larger gray region: the right corner simply corresponds to the IST amplitude  \eqref{eq:istamp} with $M^2 = 2M_{\gap}^2$ and the upper corner is close to the string amplitude with $1/\alpha' = 2M_{\gap}^2$. 
 In this sense, the lower maroon region can be thought of as the region with contributions from the spectrum  only above the cutoff $\mu_c=2$.\footnote{This is not quite true at finite $k_{\max}$, but as we increase $k_{\max}$ we find that the largest allowed coupling to a state at the gap shrinks for the points in the lower region.}
 The upper light blue/maroon islands represents the possible values of Wilson coefficients that do receive nontrivial contributions from the spin 0 and 1 states at the mass gap: these represent the non-trivial amplitudes with the hidden zero splitting constraints that do not have an infinite spin tower at the mass gap. 

\begin{figure}[t]
    \centering
    \includegraphics[width=\linewidth]{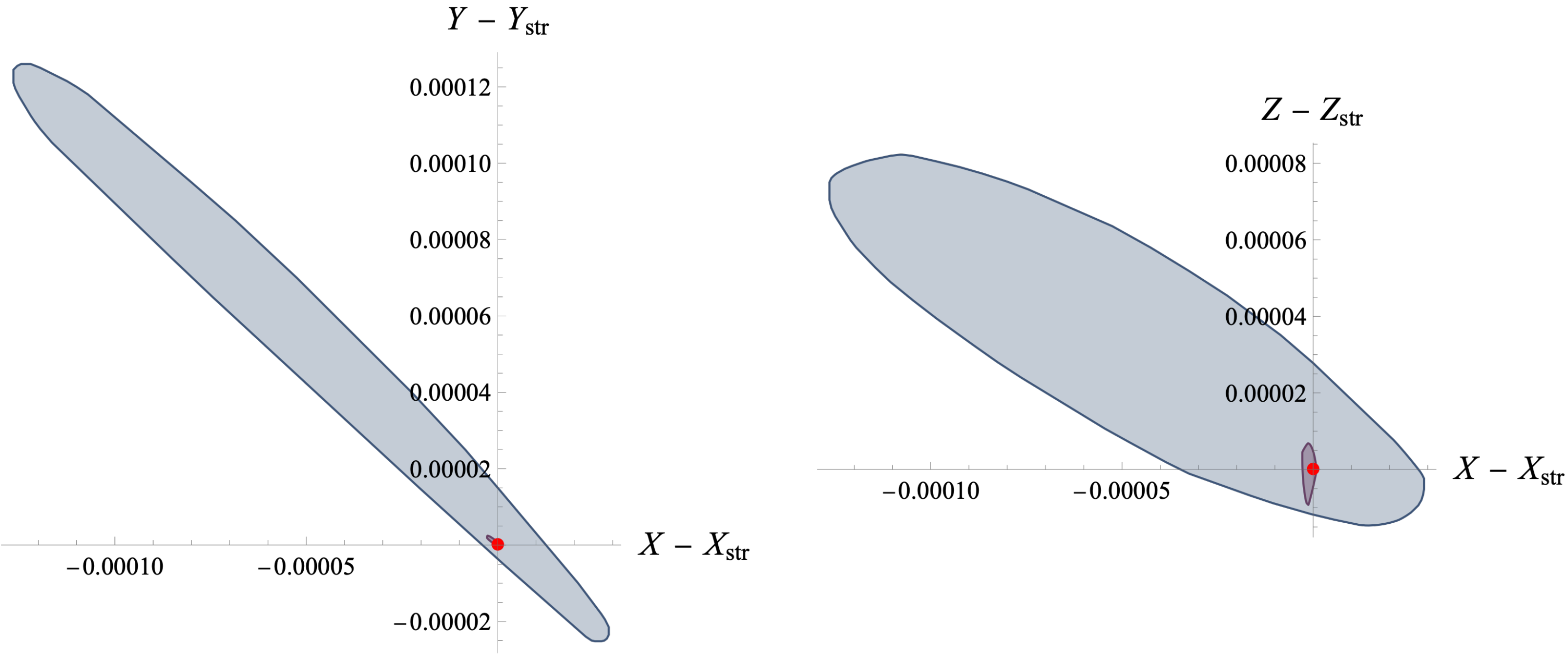}
    \caption{For  $d=10$ and $\mu_c =2$, we show the allowed $k_\text{max}=8$ island in $(X,Y,Z)$ space is projected to the $(X,Y)$ and $(X,Z)$ planes in blue. In purple, we show the same for $k_{\max} = 10$. The ratios of Wilson coefficients are shifted by the string values \reef{XYZstr} to locate the beta function (red dot) at the origin in order to better illustrate the absolute size of the islands. These plots are obtained by linearizing the nonlinear splitting constraints by scanning over both $X$ and $Y$; see Appendix \ref{app:nonlin} for details.}
    \label{fig:kmax8islandszoom}
\end{figure}

As we impose more non-linear constraints, the allowed regions shrink. Zooming in near the string amplitude, we find the $k_{\max}=6$ string island (maroon) in the bottom right of Figure \ref{fig:Zoom10d}. 
At this order in $k_{\max}$, the size of the allowed parameter space is $\mathcal{O}(10^{-2})$.

Increasing $k_{\max}$ while including the linearized constraints in \reef{nonlinAk8Yonly}, we find that the island continues to shrink. However, a more dramatic reduction in available parameter space is obtained when we include more nonlinear splitting constraints by scanning over both $X$ and $Y$. That is, we impose  \reef{k3nonlinB3} at $k=3$ instead of \reef{k3nonlinB2} and then include  both the nonlinear $k=4$ constraints in \reef{k4nonlinA}. Additionally, including the two $k=5$ constraints from \reef{nonlinA}
and the two $k=6$ constraints of \reef{nonlinAk6XY} for a total of seven nonlinear constraints (along with the standard null conditions), we find the tiny $k_\text{max}=6$ green island inside the maroon island on the bottom right of Figure \ref{fig:Zoom10d}.\footnote{To make the distinction clear, both islands on the bottom right of Figure \ref{fig:Zoom10d} were obtained with $k_{\max} = 6$, but for the maroon island we included the subset of nonlinear constraints with $k\le 6$ that can be linearized with just the $Y$-scan, whereas for the green island we scanned over both $X$ and $Y$ so that we could linearize and impose more nonlinear constraints; see Appendix \ref{app:nonlin} for further details.} Going up to $k_{\max}=8$ and imposing a total of eleven nonlinear constraints by scanning over $X$ and $Y$ both, we obtain the tiny dark blue island shown in the zoomed in plot on the left bottom of Figure \ref{fig:Zoom10d}.

Finally, by maximizing and minimizing
\be
 Z \equiv \frac{a_{3,0}}{a_{0,0}}
\ee 
 over the allowed points $(X,Y)$ in the dark blue $k_{\max}=8$ island, we obtain bounds on the range of $Z$. We  present the projection of the $(X,Y,Z)$ allowed region at $k_{\max}=8$ to the $(X,Y)$ (i.e.~zoom in on the dark blue island from Figure \ref{fig:Zoom10d}) and $(X,Z)$ in Figure \ref{fig:kmax8islandszoom}. In these plots, the Wilson coefficients are shifted by the string values
 \be
    \label{XYZstr}
    X^\text{str} 
    = \frac{\zeta_3}{\zeta_2}\,,
    ~~~~
    Y^\text{str} = \zeta_2\,,
    ~~~~
    Z^\text{str} = \frac{\zeta_5}{\zeta_2}\,,
\ee
 to center the string beta function (red dot) to the origin so that the very small scale of both islands becomes more manifest. 

In summary, the overall size of the string islands shrinks drastically as we include more non-linear constraints coming from higher $k$, so 
this gives very clear numerical evidence that, assuming there is no infinite spin tower at the mass gap, the string amplitude at $4$-points is the \textit{only} amplitude compatible with the $5$-point splitting conditions.

\subsection{Bifurcating Regions: $d=4$ and Cutoff 1.2}
\label{s:D4islands}
At the end of the previous subsection,  we made a strong claim: that the string was the only amplitude compatible with the splitting constraints that did not have an infinite tower of spinning states at the same mass. However, we only gave evidence for this in the string-inspired case of $d=10$ and cutoff $\mu_c=2$. 
The general claim  requires investigating the bounds for different $d \le 10$ and $\mu_c< 2$. 
In order to directly rule out the amplitudes with spin towers, we still need to take $\mu_c > 1$ such that only scalars/vectors can be exchanged at $s=M_{\gap}^2$.

\begin{figure}[t]
    \centering
    \includegraphics[width=0.55\linewidth]{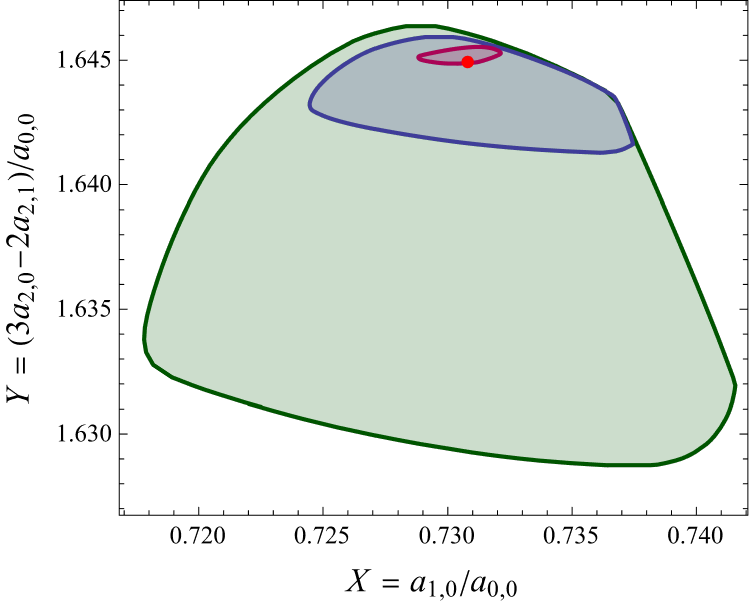}
    \caption{Using the $Y$-scan constraints \reef{nonlinAk8Yonly} and assuming no spin-towers at the mass-gap, the $d=10$ string-island allowed regions in $(X,Y)$-plane are shown for  $k_\text{max}=8$ and cutoff $\mu_c = 1.2$ (green), $1.5$  (blue), and $2$ (purple). As we decrease the cutoff at fixed $k_{\max}$, the island gradually increases in size.}
    \label{fig:kmax8cutoffislands}
\end{figure}

Varying $\mu_c$, we find that the allowed region still bifurcates at sufficiently high $k_\text{max}$. 
The amplitude is allowed to have any (positive) spectrum in the range $(\mu_c,\infty)$, so 
lower values of $\mu_c$ generally give weaker bounds. This is illustrated in Figure \ref{fig:kmax8cutoffislands}, 
which compares the sizes of the 
$d=10$, $k_{\text{max}}=8$ islands for $\mu_c = 1.2$ (green), $1.5$ (blue) and $2$ (purple).
While the island size and details of the bounds at given finite $k_{\max}$ depend on the choice of $d$ and $1< \mu_c \le 2$, the appearance of the island and its continued shrinking happens regardless of their particular values in all the cases we have tested.

\begin{figure}[t]
    \centering
    \includegraphics[width=\linewidth]{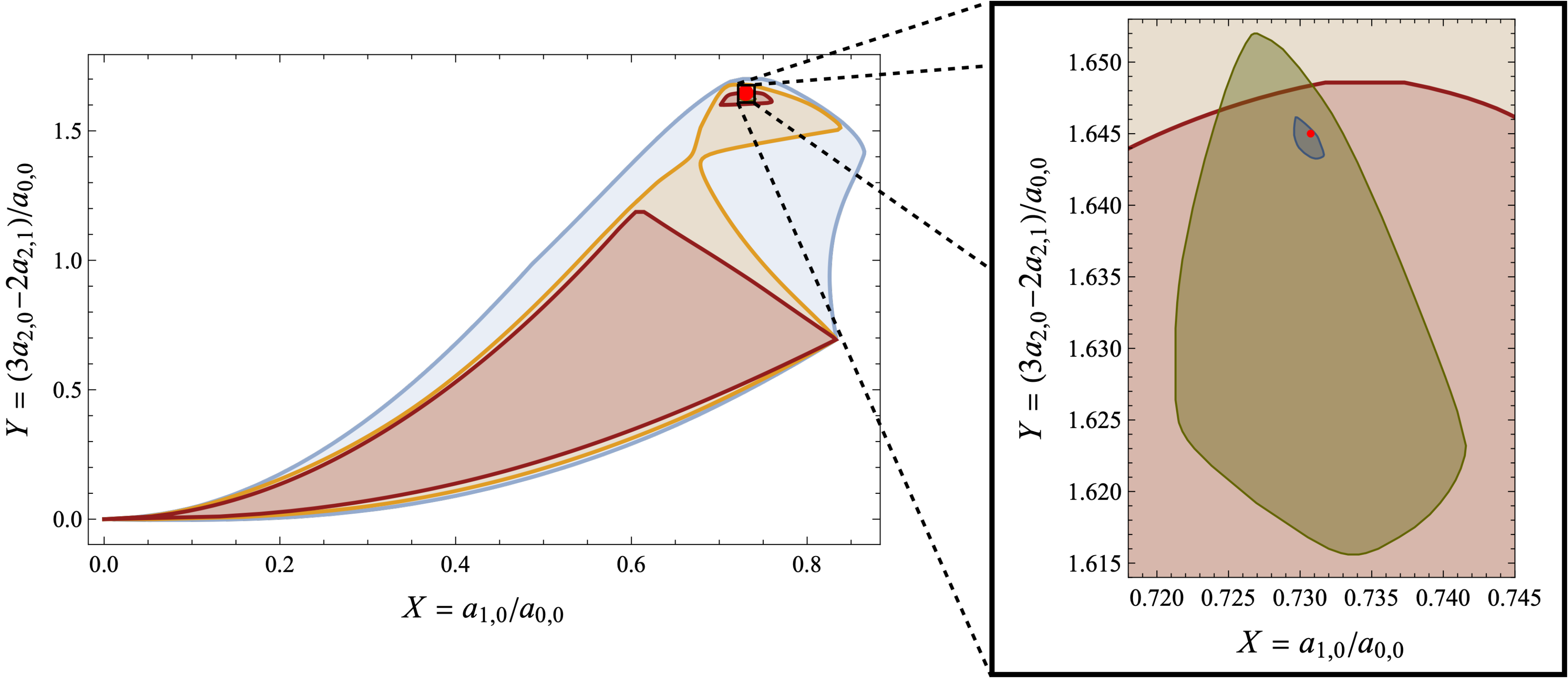}
    \caption{Allowed regions in the $(X,Y)$-plane for $d=4$ and cutoff $\mu_c=1.2$ assuming only a finite number of states at the mass gap. 
    {\bf Left:} At $k_{\max} = 4$ (blue) and $5$ (orange) with $Y$-scan constraints \reef{nonlinAk8Yonly}, the region has not yet bifurcated, but it does at $k_\text{max}=8$ (maroon). 
    {\bf Right:} Zoom-in near the string amplitude with $\alpha'M_{\gap}^2 = 1$ (red dot). The green island is for $k_{\max} = 6$ with the seven $k \le 6$ nonlinear constraints available with a double-scan over $X$ and $Y$, while the dark blue is the same for $k_{\max} = 8$ and the eleven nonlinear constraints with $k\leq 8$.}
    \label{fig:Zoom4d}
\end{figure}

To further illustrate this point, we pick $d$ as far from criticality as possible while still maintaining unitarity and sufficiently generic kinematics to be able to impose the 5-point splitting conditions: this is $d=4$. Then taking the cutoff just slightly above the mass gap, we set $\mu_c = 1.2$.   
Figure \ref{fig:Zoom4d} shows the resulting allowed regions using the nonlinear constraints available with scanning over $Y$. At $k_\text{max} = 4$ and $5$, the regions (blue and orange) have  not bifurcated,  so there is no clear separation between theories with and without the scalar at the mass gap. 
With increasing $k_{\max}$, the region becomes highly concave and eventually the ``bridge'' between the two regions breaks. At $k_\text{max} = 8$ (maroon region) the region has bifurcated. As in the previous section, the lower part of the bifurcated region is simply a scaled down version of the general region, but now with mass gap scaled to $\mu_c = 1.2$. The island around the string is quite a bit bigger than for the $d=10$ and $\mu_c = 2$ case in the previous subsection. However, as we include additional nonlinear constraints by scanning over both $X$ and $Y$, the island shrinks significantly, as shown for $k_\text{max}=6$ (green) and $k_\text{max}=8$ (dark blue) in the zoom-in plot in Figure \ref{fig:Zoom4d}. 

While the $k_{\max}=8$ island for $d=4$ and $\mu_c = 1.2$ (dark blue region) is not as small as in the equivalent case of $d=10$ and $\mu_c = 2$ (dark blue island on the bottom left of Figure \ref{fig:Zoom10d}), it is still nontrivial evidence that the island  continues to shrink to a point as $k_{\max}$ increases, leaving the string amplitude as the only solution. Testing other values of the cutoff gives the same result: for large enough $k_{\max}$, the region bifurcates and the island with contributions from the scalar/vector at the gap shrinks around the string coefficients.

\subsection{Bounds on the Wilson Coefficients}
\label{sec:bounds}

Let us compare the results in Sections \ref{s:D10islands} and \ref{s:D4islands} by converting the islands to bounds on the  lowest available Wilson coefficients. 

Using the $k_\text{max}=8$ and $10$ islands for the case of $d=10$ and $\mu_c =2$ and the  $k_\text{max}=8$ island for $d=4$ and $\mu_c =1.2$, we obtain the bounds listed in Table \ref{tab:bounds} on the ratios
\be
    X = \frac{a_{1,0}}{a_{0,0}}
    \,, ~~~~
    Y = \frac{3a_{2,0}-2a_{2,1}}{a_{0,0}} 
     \, ,~~~~
   Z =\frac{a_{3,0}}{a_{1,0}} \,.
\ee
It is expected, the bounds in the critical case $d=10$ are stronger than in the sub-critical dimension $d=4$. However, even in the $d=4$ case with $\mu_c=1.2$ far from the next stringy mass state at 2, the island continues to shrink as $k_\text{max}$ increases and already at $k_\text{max}=8$ it leaves only a interval of less than $2.9 \cdot 10^{-3}$ for $X$, $Y$, and $Z$. To compare, for $d=10$ with  $\mu_c=2$, the biggest island constrains range of  $X$, $Y$, and $Z$ to less than $1.7\cdot 10^{-4}$   for $k_\text{max}=8$ while for $k_\text{max}=10$ it shrinks to less than $4\cdot 10^{-6}$. This leaves very little parameter space for anything not the string beta function amplitude.
\begin{table}[t]
\setlength{\tabcolsep}{6pt}
\renewcommand{\arraystretch}{1.2}
\begin{tabular}{c|c|c|c}
\multicolumn{4}{c}{$d=10$, $\mu_c=2$} \\
 & min-max $a_{1,0}/a_{0,0}$ 
 & min-max $(3a_{2,0} - 2a_{2,1})/a_{0,0}$ 
 & min-max $a_{3,0}/a_{0,0}$ \\ \hline
$k_{\max}=8$ 
    & 0.73063 - 0.73080       
    & 1.64490 - 1.64507          
    & 0.63036 - 0.63046       \\ 
$k_{\max}=10$ 
    & 0.730760 - 0.730764       
    & 1.644933 - 1.644937        
    & 0.630367 - 0.630384       \\ \hline \hline
String Value 
    & 0.730763\ldots 
    & 1.644934\ldots           
    & 0.630376\ldots  \\
\noalign{\vskip 3mm}
\multicolumn{4}{c}{$d=4$, $\mu_c=1.2$} \\
 & min-max $a_{1,0}/a_{0,0}$ 
 & min-max $(3a_{2,0} - 2a_{2,1})/a_{0,0}$ 
 & min-max $a_{3,0}/a_{0,0}$ \\ \hline
$k_{\max}=8$ 
    & 0.72958 - 0.73167       
    & 1.64326 - 1.64614 
    & 0.62955 - 0.63185 \\ \hline \hline
String Value 
    & 0.730763\ldots 
    & 1.644934\ldots           
    & 0.630376\ldots       \\
\end{tabular}

\caption{\label{tab:bounds}Table of bounds on a selection of lowest Wilson coefficients. The bounds are extracted from the island plots in Figures \ref{fig:kmax8islandszoom} and \ref{fig:Zoom4d}. 
The string values are obtained from the exact values \reef{XYZstr}.}
\end{table}

Hence, based on the numerical results, it is natural to conjecture that in any dimension $4 \le d \le 10$ the string amplitude is the only unitary solution consistent with the hidden zero and the $5$-point splits so long as there are only finitely many states at the mass gap $M_\text{gap}$. Every other unitary amplitude must then either have a spin tower at $M_\text{gap}$ or only contributions above the cutoff $\mu_c$. For a theory with only contributions above the cutoff, we can do the same kind of analysis to show that either it must have a spin tower at the cutoff or it is the string amplitude with $\mu_c = 1/(\alpha'M_{\gap}^2) > 1$. Therefore, even for amplitudes without contributions at the mass gap, we conclude that the only theory without an infinite spin tower with all states of the same mass is the string amplitude.

\section{Outlook}
\label{sec:outlook}
The goal of this paper was to study the impact of constraints from  higher-point amplitudes on the bootstrap of 4-point scattering processes. We have worked with the example of EFTs whose amplitudes satisfy the hidden zero and split constraints \cite{Arkani-Hamed:2023swr} and shown how the constraints from 5-point splits carve out a non-convex subregion in the parameter space of allowed values of the Wilson coefficients. This contrasts the convexity of the allowed  regions obtained from purely 4-point positivity considerations. We consider this a proto-type example that may well have counterparts in other contexts, such as for supersymmetric theories, effective theories of Goldstone modes, etc. Since the general higher-point unitarity bootstrap is a challenging open problem, this basic way of extracting constraints on the 4-point amplitude from fundamental factorization properties of the higher-point amplitudes can give valuable insights into how additional parameter space is ruled out in the bootstrap. 

We also demonstrated that when  infinite spin towers at the mass gap $M_\text{gap}$ are excluded, the allowed region bifurcates into separate regions, one of which we found to be a shrinking island around the beta function string amplitude \reef{eq:4ptStringAmp}. We presented results for both the string-inspired case of $d=10$ and cutoff $\mu_c =2$ and the much more generic case of $d=4$ and cutoff $\mu_c =1.2$. In both cases, the string island continues to shrink as more constraints from higher-derivative terms are included and in the former case, we find three of the lowest Wilson coefficients fixed to within $4 \cdot 10^{-6}$ of the string values at order $k_\text{max}=10$. 

\vspace{2mm}
\noindent {\bf Comparisons with other recent perturbative string bootstraps.} 

\begin{figure}[t]
    \centering
    \includegraphics[width=0.7\linewidth]{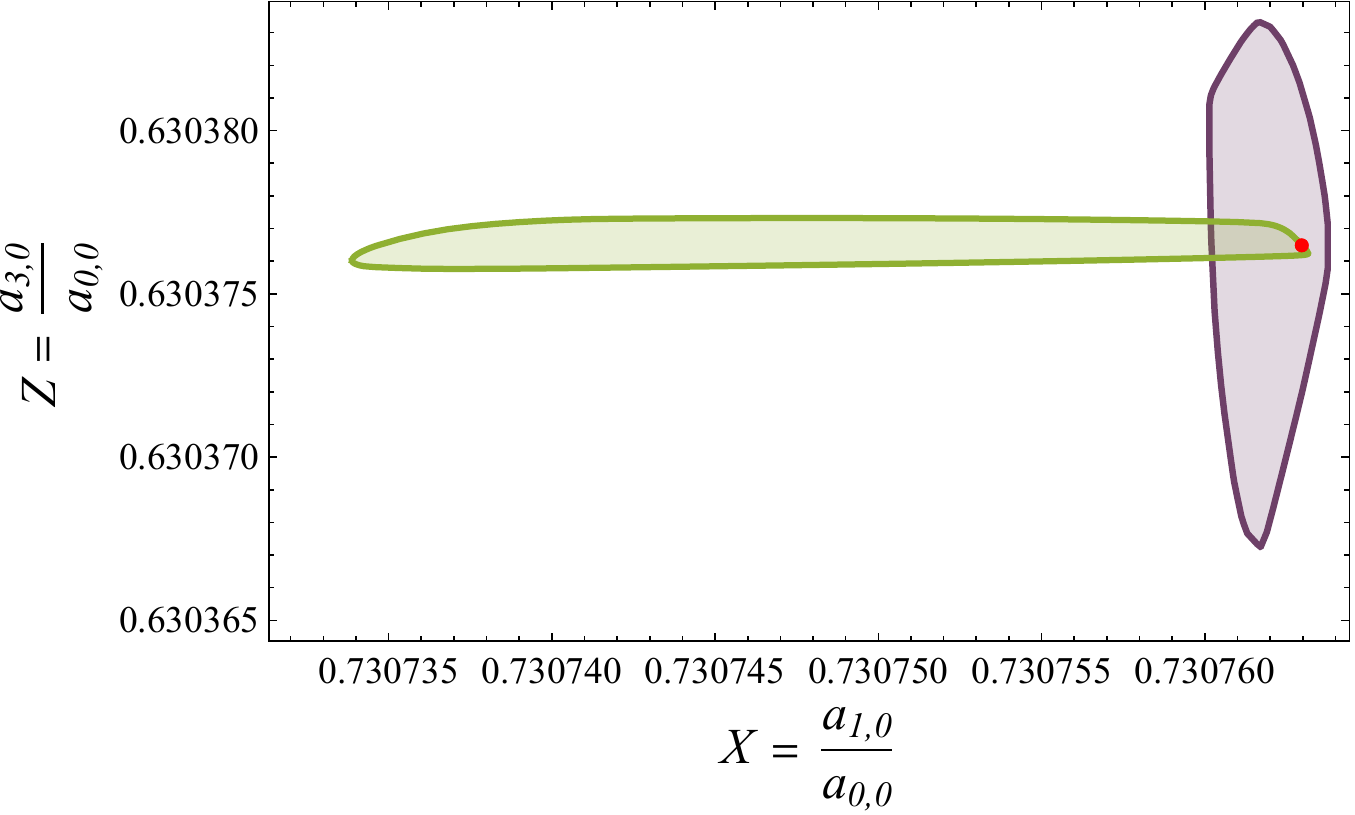}
    \caption{Comparison on the $D=10$ islands in the $(X,Z)=(a_{1,0}/a_{0,0},a_{3,0}/a_{0,0})$ obtained in two different contexts: in purple is the tiny $k_{\max}=10$ island from the righthand side of Figure \ref{fig:kmax8islandszoom} obtained with the hidden-zero nonlinear constraints and assuming no infinite spin tower at the mass gap. The green island is obtained by assuming the amplitude to be maximally supersymmetric and obeying the string monodromy relations. It is obtained with $k_{\max}=8$. The string amplitude (red dot) lies inside both islands and very close to the bound of the monodromy island. 
    } 
    \label{fig:monovssplit}
\end{figure}
The size of our $k_\text{max}=10$ island for 
$d=10$ and cutoff $\mu_c =2$ is comparable to the island obtained in a very different bootstrap analysis \cite{Huang:2020nqy,Berman:2024wyt,Chiang:2023quf} in which string monodromy relations where used as additional input. Specifically, ref.~\cite{Berman:2024wyt} considered 
the 4-point amplitude of external complex scalars associated with the massless $\mathcal{N}=4$ super Yang-Mills  supermultiplet and parameterized it as
\be
\mathcal{A}_4^{\text{SUSY}}[z_1z_2\bar{z}_3\bar{z}_4]
=
s^2
\tilde{\mathcal{A}}(s,u)  \, 
\ee
with $\tilde{\mathcal{A}}_4(u,s)=\tilde{\mathcal{A}}_4(s,u)$. 
The low-energy expansion of this 
$\tilde{\mathcal{A}}$ takes exactly the same form as  $\tilde{\mathcal{A}}_4(s,u)$ in this paper, with our $\mathcal{A}_4(s,u) = (s+u)\tilde{\mathcal{A}}_4(s,u)$; see \reef{AtildeDef}-\reef{A4Contact}. We can therefore directly compare the
bounds on the $a_{k,q}$ coefficients in the two cases. The two $d=10$ string islands are shown in Figure \ref{fig:monovssplit}. 
The key assumption in 
\cite{Huang:2020nqy,Berman:2024wyt,Chiang:2023quf}
is that the 4-point $\mathcal{N}=4$ SYM EFT amplitude 
obeys the (low-energy expansion of) the string monodromy relations \cite{Plahte:1970wy,Stieberger:2009hq,Bjerrum-Bohr:2009ulz,Bjerrum-Bohr:2010mia,Bjerrum-Bohr:2010pnr} that relate tree-level open string amplitudes with different orderings of the massless external states. This is a much more severe restriction, for example in that  it fixes some of the  Wilson coefficients directly to their string values. Moreover, since the string monodromy relations arise from writing the string amplitude as a worldsheet integral, it is very much a string-theoretic input. In contrast, in the analysis performed in this paper we use a purely field theoretic input: the hidden zeros and splits are already features of the $\Tr[\Phi^3]$ amplitudes and require no reference to any string-y information. Therefore, the fact that the string amplitude appears to be uniquely picked out by the splitting conditions and unitarity as the only compatible amplitude with no infinite towers of spinning states is more surprising than that the string amplitude was isolated by assuming by monodromy + unitarity.

Another bottom-up bootstrap of the Veneziano amplitude was considered recently in \cite{Berman:2024wyt} and \cite{Albert:2024yap}. Here constraints on the low-mass spectrum was considered for massless scalar amplitudes in $\mathcal{N}=4$ super Yang-Mills. In that case, using zero-subtracted dispersion relations, it was shown that if the spectrum had a scalar at the mass gap and a cutoff $\mu_c$ to the next possible state, there was a 1-parameter family of new corner theories parameterized by $\mu_c$; they included the massless Veneziano amplitude for $\mu_c = 2$. With $d=10$ and $\mu_c = 2$, choosing the cubic coupling of the massive scalar to the massless states to be the string value (a much milder input than the monodromy relations), the allowed space of Wilson coefficients shrank to an island around the string. In that case, as well as here, the island continued to shrink with increasing $k_\text{max}$. The best bound obtained in \cite{Berman:2024wyt} for the equivalent of $X = a_{1,0}/a_{0,0}$ was 
$0.7261 < X  < 0.7333$ for $k_\text{max}=18$. In contrast, using the nonlinear relations from the 5-point split constraints, we found in this paper  
$0.730760 < X  < 0.730764$ for $k_\text{max}=10$. 

\vspace{2mm}
\noindent {\bf Island Spectrum Extraction.} After using SDPB to minimize or maximize a coefficient, it is possible to extract the spectrum of the extremal solution to the constraints. In \cite{Albert:2023seb}, for example, it was found that extracting the amplitude living near a kink in their bounds on large-$N$ QCD had a spectrum quite similar to that of real world pion scattering. However, one common feature of these extremal spectra is that the theories living near the boundaries appear to have only a single Regge trajectory \cite{Caron-Huot:2021rmr,Albert:2023seb,Berman:2024wyt,Albert:2024yap}. This is in direct contrast with string theory, which has an infinite set of daughter trajectories below the leading trajectory. Further, the numerical spectra often have many spurious states above the leading trajectory (i.e.~at high spin and low mass), though these typically have small couplings relative to the states on the leading trajectory and such states can be analytically shown to be disallowed at infinite $k_{\max}$ \cite{Berman:2024kdh,Berman:2024owc}. It is known that such single trajectory amplitudes are inconsistent \cite{Eckner:2024pqt}, so their appearance in the numerical bootstrap is likely due to the fact that theories with daughter trajectories only appearing at infinite mass are allowed and so single trajectory theories cannot be disallowed at finite truncation order.

\begin{figure}[t]
    \centering
    \includegraphics[width=0.7\linewidth]{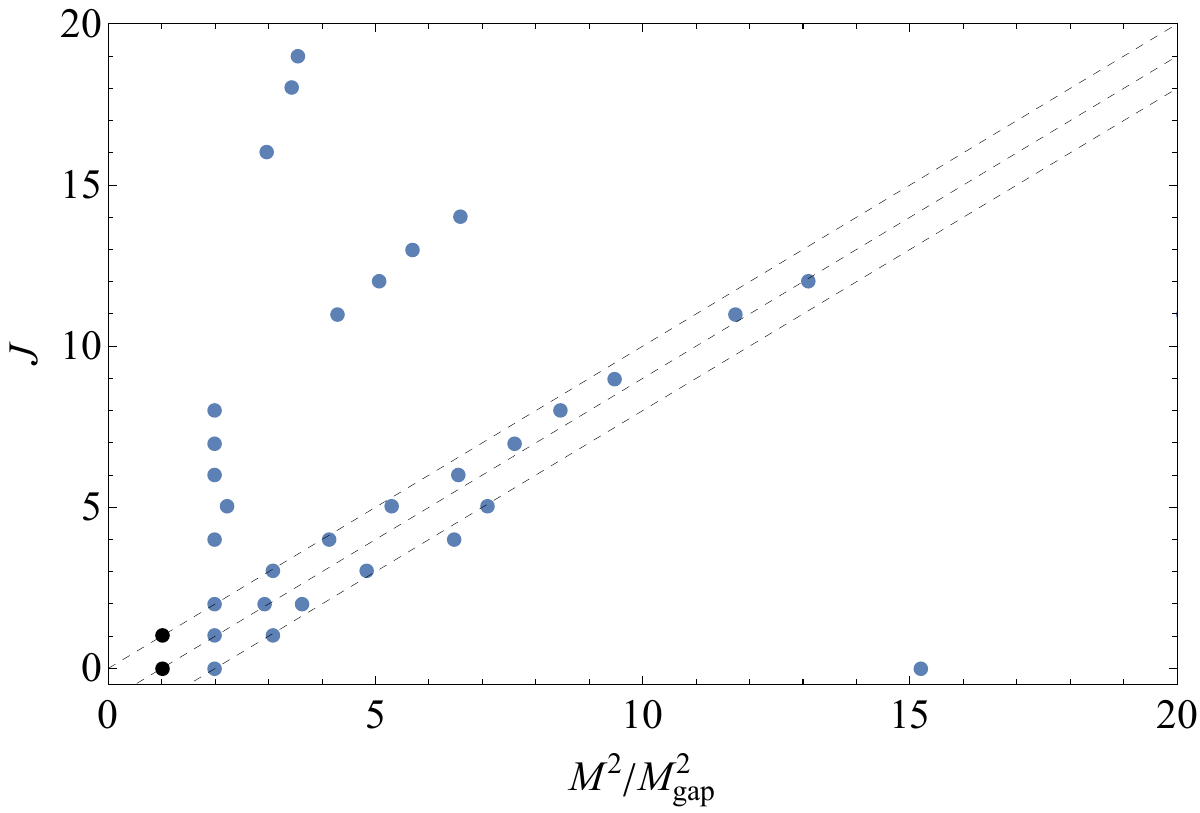}
    \caption{Spectrum for the amplitude with minimal $a_{3,0}/a_{0,0}$ in $d = 10$, a scalar and vector at the gap, and cutoff $\mu_c = 2$ with $X$ and $Y$ fixed to their string values and all double-scan constraints imposed up to $k_{\max} = 10$. In black are the spin zero and one states which we input by hand, while blue points correspond to the spectrum found by SDPB. The gray dashed lines correspond to the first three trajectories of the string amplitude.}
    \label{fig:d10spec}
\end{figure}

It was shown in  \cite{Chiang:2023bst} that in the bootstrap using monodromy constraints, the numerical spectrum near the string amplitude gets a few of the states on daughter trajectories, though still with many spurious states. It is interesting to consider whether the split constraints considered in our paper can improve the numerical spectrum to be more comparable to that of the string amplitude. Figure \ref{fig:d10spec} shows the spectrum extracted for minimizing $a_{3,0}/a_{0,0}$ in $d = 10$ with cutoff $\mu_c = 2$ and $X$ and $Y$ fixed to the string values to linearize as many as possible of the nonlinear split constraints in the double-scan approach. We still see some spurious states living above the leading Regge trajectory, but we additionally find a few states living on or near the first two daughter trajectories! It would be interesting to continue fixing coefficients such that \textit{all} nonlinear constraints up to $k_{\max} = 11$ can be imposed to evaluate whether these constraints can help better match the spectrum of string theory rather than just the numerical Wilson coefficients. Likewise, the spectrum may improve if the analytic results \cite{Berman:2024kdh,Berman:2024owc} could be implemented to avoid the states above the leading Regge trajectories, for example via the stronger assumption of a slope 1 leading Regge trajectory  as in \cite{Haring:2023zwu,Berman:2024wyt}. 

\vspace{2mm}
\noindent {\bf Harder Regge behavior.} For the purpose of illustrating the consequences of higher-point constraints on the 4-point amplitude, we assume in our bootstrap analysis that the 4-point amplitude has the improved Regge behavior \reef{eq:bounduandtfixed} rather than large $|s|$ behavior of the standard Froissart bound
\be
  \lim_{|s| \to \infty} \frac{\mathcal{A}_4(s,u)}{s^2} =0 \,,~~~~~
  \lim_{|s| \to \infty} \frac{\mathcal{A}_4(s,-s-t)}{s^2} =0 \,.
  \label{eq:bounduandtfixed2}
\ee
Relaxing our Regge assumption to \reef{eq:bounduandtfixed2}, the story is less sharp. 
With the now twice-subtracted dispersion relations, the allowed region of the Wilson coefficients $a_{k,q}/a_{1,0}$ can certainly still be seen to be non-convex. But when assuming no infinite tower of states at the mass gap, the allowed parameter space cannot cleanly be shown to bifurcate into parts with and without particle content at the mass gap. We give a more detailed analysis of this scenario in Appendix \ref{app:2sdr}, but one major difference with the improved Regge scenario is that there is no analogue of the single scan: in order to impose \textit{any} of the nonlinear constraints, we have to scan over two variables. 
Because the allowed region becomes very thin and narrow with increasing $k_\text{max}$, 
the essential problem is that we can not guarantee that bifurcation occurs; the island is so thin that it could ``escape'' between probed points in the finite grid. Employing the more powerful Skydive tool \cite{Liu:2023elz}, developed for the conformal bootstrap,  may improve the analysis and possibly isolate the string also the case with generic large $|s|$ behavior.

\vspace{2mm}
\noindent {\bf Constraints from and at higher-points.}
Let us now turn from the numerical bounds to the analytic relationship between Wilson coefficients imposed by the hidden zeros and splits. We showed in Section \ref{sec:HZ} that the 5-point splits \reef{eq:5ptSplitA}-\reef{eq:5ptSplitB} fix nonlinear relationships among the 4-point Wilson coefficients $a_{k,q}$. Specifically, at each $k$ (corresponding to the $k+1$ degree Mandelstam polynomial in $\mathcal{A}_4$ and hence $(2k+2)$-derivative order in the adjoint scalar EFT Lagrangian) there are ($\lfloor  k/2 \rfloor$+1) free coefficients $a_{k,q}$, labelled by $q=0,1,2,\lfloor  k/2 \rfloor$, once crossing symmetry 
$\mathcal{A}_4(u,s) = \mathcal{A}_4(s,u)$ and the hidden zero condition $\mathcal{A}_4(s,-s) = 0$ have been imposed. 

It is straightforward to construct general EFT 5-point amplitudes $\mathcal{A}_5$ that factorize correctly to $\mathcal{A}_3 \times \mathcal{A}_4$ on the massless  2-particle poles. The residues of $\mathcal{A}_5$ are determined by the $a_{k,q}$ and the cubic coupling of Tr($\Phi^3$) and the 4-point Wilson coefficients $a_{k,q}$. 
Local terms are cyclic polynomials in the Mandelstams and can have arbitrary coefficients. This construction is valid for any choices of the $a_{k,q}$ and is highly non-unique due to the large number of possible local terms. A key point in this paper is that this may no longer be the case when specific properties are required of the amplitudes. 

Once we impose the hidden zero conditions on $\mathcal{A}_5$, many coefficients of the local terms are fixed, however, the $a_{k,q}$ remain unconstrained. 
As we have shown in Section \ref{sec:HZ}, this changes the moment we also impose the 
5-point splits \reef{eq:5ptSplitA}-\reef{eq:5ptSplitB}: many more local terms are fixed then, but more importantly now nonlinear constraints relate the $a_{k,q}$. In fact, we found that each $a_{k,q}$ with $q \ge 1$ is fixed via these nonlinear relation in terms of the cubic $\Tr(\Phi^3)$ coupling $g$, lower-order Wilson coefficients $a_{k',q'}$ with $k'<k$, and $a_{k,0}$. (See Table \ref{tab:Constraintsk6}.) This leaves as the only free coefficient $a_{k,0}$ for each values of $k = 1,2,\ldots$. This means that, surprisingly, there is only a single free Wilson coefficient of the 4-field local operators at each derivative order in the EFT expansion.  

Extending the analysis to 6-points (see Appendix \ref{app:6ptsplit}), we find no additional constraints among the 4-point coefficients, but in principle that does not exclude that 7- or higher-point splits could yield additional  constraints among the $a_{k,q}$. However, since we know the string amplitudes are compatible with the splits \cite{Arkani-Hamed:2023jwn}, any resulting nonlinear constraints also have to be solved by the coefficients of its low-energy expansion. It is easy to see for the string beta function amplitude that $a_{k,0} = \zeta_{k+2}$ and  that there are only simple zeta values of depth 1 (i.e.~no $\zeta_{3,5}$ etc). Hence for even $k$, $a_{k,0} \propto \pi^{k+2}$ and it is conceivable that these Wilson coefficients can be further fixed from higher-point splits. However, the $\zeta_\text{odd}$ values are not expected to satisfy any algebraic relations with other $\zeta_n$'s, so for $k$ odd, $a_{k,0}$ can never be fixed by any split conditions. 

Another interesting feature of the analysis in Section \ref{sec:HZ} (and with further details in Appendix \ref{app:LocalTerms}) is that 
a large number of the local terms in the 5- and 6-point amplitudes are fixed in terms of the lower-point Wilson coefficients when we impose the splits. Going up to $k=8$ (18 derivative order), the general ansatz \reef{5ptansatz} for the 5-point amplitude to this order has 259 free coefficients in the contact terms. However, the 5- and 6-point splits fix all but 12 of those (see Table \ref{tab:LocalTerms} in Appendix \ref{app:LocalTerms}). Again, it is possible that imposing splits at higher-points would fix additional terms; though never coefficients that the string tree amplitudes would predict not to satisfy algebraic relations, such as those that are multiple zeta-values. We discussed aspects of this in Section \ref{app:MVZ}. 

Let us now turn to the question of what the nonlinear relationships among the $a_{k,q}$ imply for the structure of the 4-point amplitude.

\vspace{1mm}
\noindent {\bf Factorization and Exponentiation.}
Examples of how symmetries imposed at higher-point constrain the Wilson coefficients of the 4-point low-energy expansion have been explored previously in other contexts. For example, it was shown that imposing the linear constraints of the KK \cite{Kleiss:1988ne} and BCJ \cite{Bern:2008qj} relations at 6-point implies  nonlinear constraints among the 4-point Wilson coefficients of bi-adjoint \cite{Chen:2022shl} and adjoint scalar amplitudes \cite{Brown:2023srz}. In the bi-adjoint case \cite{Chen:2022shl} it turns out that the  nonlinear constraints lead to a factorization of the 4-point amplitude into a function that depends only on the lowest-order Wilson coefficient and an exponentiated factor symmetric in $s,t,u$. In our case here, with nonlinear constraints arising from the hidden  splits, it also turns out that 
the 4-point amplitude factorizes and that one factor exponentiates! To see this, note that upon imposing the nonlinear constraints in Table \ref{tab:Constraintsk6} (and beyond for higher $k$), $\mathcal{A}_4$ depends only on the Wilson coefficients $a_{k,0}$. To motivate the factorized form of $\mathcal{A}_4$, recall that for the beta function amplitude, $a_{k,0} = \zeta_k$, and hence for even-$k$, the $a_{k,0}$ are all associated with powers of $\pi$ while the odd-$k$ coefficients are algebraically independent. Motivated by this, let us separate the  even and odd $k$ coefficients in the  EFT amplitude with general $a_{k,0}$'s by defining
\be
  f_0(s,u) = \mathcal{A}_4(s,u) \bigg|_{a_{2k+1} = 0}
  = -\frac{g^2}{s u}
  + a_{0,0}
  -  \frac{1}{2g^2}a_{0,0}^2 s u
  + \Big(s^2 + \frac{3}{2}su + u^2\Big)a_{2,0}
  + \ldots
\ee
We can then write 
\be
  \mathcal{A}_4(s,u) 
  = f_0(s,u) \,e^{V}\,,
\ee
where we determine $V$ to be a fully $stu$-symmetric polynomial that takes the form
\be
  \begin{split}
     V(s,t,u)=& 
     \frac{stu}{g^2}
     \bigg[
       a_{1,0}
       + \frac{1}{2}(s^2+t^2+u^2) a_{3,0}
       +\frac{1}{2^2}(s^2+t^2+u^2)^2 a_{5,0}
       \\
       & \hspace{10mm}
       + \bigg( \frac{1}{2^3}(s^2+t^2+u^2)^3 
       +\frac{1}{3} (s t u)^2 \bigg) a_{7,0}
       \\
       & \hspace{10mm}
       + \bigg( \frac{1}{2^4}(s^2+t^2+u^2)^4 
       +\frac{1}{2} (s t u)^2(s^2+t^2+u^2) \bigg) a_{9,0}
       \\
       & \hspace{10mm}
       + \bigg( \frac{1}{2^5}(s^2+t^2+u^2)^5 
       +\frac{1}{2} (s t u)^2(s^2+t^2+u^2)^2 \bigg) a_{11,0}
       \\
       & \hspace{10mm}
       + \bigg( \frac{1}{2^6}(s^2+t^2+u^2)^6 
       +\frac{5}{12} (s t u)^2(s^2+t^2+u^2)^3
       +\frac{1}{5} (s t u)^4
       \bigg) a_{13,0}
       + \ldots
     \bigg]
  \end{split}
\ee
We include the high orders in the expansion to show that new symmetric polynomial structures  continue to appear in at higher orders. In the case of the bi-adjoint scalar, the factorization was closely related to the fact that the nonlinear constraints implied the monodromy relations. In our case here, it is less clear if there is such a resulting structure. Perhaps putative constraints from higher-point splits would be needed.

\vspace{1mm}
\noindent {\bf Bounds on higher-point Wilson coefficients.}
It was recently shown that, under particular assumptions about the analyticity and polynomial boundedness, there are moment-type inequalities among subsets of coefficients for local terms of $n$-point amplitudes for arbitrary $n$ \cite{Cheung:2025krg}. In particular, the simplest example of these constraints relates particular coefficients at $4$, $5$, and $6$ point:
\be\label{eq:cgcond}
    a_{k-1,0}z_{3k}-w_{2k,1}^2 \geq 0\, ,
\ee
where $w_{2k,1}$ is the coefficient of $(X_{1,3}X_{1,4})^k$ in \eqref{eq:5ptLoc} and $z_{3k}$ is the coefficient of $(X_{1,3}X_{1,4}X_{1,5})^k$ in the contact piece of the 6-point amplitude described in Appendix \ref{app:6ptsplit}. 

One could imagine that, if $z_{3k}$ and $w_{2k,1}$ are completely fixed in terms of $a_{k,q}$ coefficients by the splitting constraints, the inequality \reef{eq:cgcond} could provide yet another constraint on the $4$-point coefficients. However, we find that the splitting conditions trivialize \reef{eq:cgcond} at least for the low-$k$ examples we have tested. For example, at $k = 1$, the splitting conditions require
\begin{equation}
    w_{2,1} = \frac{a_{0,0}^2}{g} 
    ~~~\text{and}~~~
    z_{3} = \frac{a_{0,0}^3}{g^2}\,
\end{equation}
so that \eqref{eq:cgcond} is exactly saturated. At $k = 2$, we have $w_{4,1} = \frac{a_{1,0}^2}{g}$ and $z_{6} = \frac{a_{1,0}^3}{g^2}$, so, again, the bound is saturated by the split conditions. Therefore, these bounds do not appear to yield any additional constraints in our context.

Since the splitting conditions yield coefficients that saturate the bounds such as \reef{eq:cgcond}, one might ask if the argument can be reversed: can assuming saturation of the bounds like \reef{eq:cgcond} replace the splitting constraints? Table \ref{tab:LocalTerms} gives the large number of possible contact terms in the ansatz for the 5- and 6-point amplitudes and that has to be contrasted with the fact that bounds such as \reef{eq:cgcond} only involve the coefficient of very few of these terms. 
Therefore, far fewer coefficients could be fixed by requiring the saturation of bounds such as \reef{eq:cgcond} than can be by using the splits. Further, one would not recover the constraints among 4-point coefficients.

\vspace{1mm}
\noindent {\bf Splits and massive factorizations at higher points.} While a higher-point EFT amplitude can be built up systematically from lower-point amplitudes via factorization on the massless poles, it is not possible to check from the bottom-up EFT approach whether these objects correctly factorize on the massive residues. Along these lines, in \cite{Arkani-Hamed:2023jwn}, it was shown that it is trivial to build higher-point deformations of string amplitudes that correctly factorize on massless poles, that however fail to have consistent factorization on \textit{massive} residues. Quite remarkably, one can easily check that these deformations also fail to have the split factorizations studied in this paper. One natural question is then whether there is any conceptual connection between the massless split factorizations and correct behavior on massive poles, explaining why these two different behaviors seem to be comparably constraining.

\vspace{1mm}
\noindent {\bf Going beyond scalar theories.} 
While the splits and the associated hidden zeros do not as yet have a simple, conventional understanding in field theory, they do have a fundamental origin in the ``surfaceology'' formulation of amplitudes, not just at tree-level but at all loop orders, and extend beyond toy scalar theories to a wide range theories of real-world interest, from pions to gluons, and also for some classes of uncolored theories.  It would therefore be interesting to extend our bootstrap analysis to these more interesting theories as well. The application to gluon amplitudes is especially interesting --- as a first step, the dispersive representation of gluon amplitudes, involving the spinning Gegenbauer polynomials --- should be translated into the ``scalar scaffolding'' formalism \cite{Arkani-Hamed:2023jry}, a step of independent technical interest in its own right. 

\vspace{1mm}
Beyond the splitting conditions, the idea of using higher-point information to constrain 4-point EFT amplitudes in conjunction with the bootstrap conditions is underexplored. Essentially, any behavior beyond standard factorization that relates higher point amplitudes to lower point could generate new conditions on 4-point amplitudes. For example, one could consider whether there exist constraints from the soft bootstrap \cite{Elvang:2018dco}  on 4-point Wilson coefficients which can be leveraged to isolate unique amplitudes.  A study of constraints from supersymmetry at higher point
is under way. 
If such constraints were to exist, they could have similar nonlinear properties to those found from the splittings, in which case a method for linearization of the constraints would be necessary to input them into the bootstrap.

\acknowledgments 
The authors thank Jan Albert, Nima Arkani-Hamed, Brando Bellazzini, David Berenstein, Aditi Chandra, Johan Henriksson, Kelian H\"aring, Nick Geiser, David Gross, Andrea Guerrieri, Denis Karateev, Loki Lin, Roger Morales, and João Penedones for useful discussions. We also thank Oliver Schlotterer for a detailed explanation of the appearance of MZVs in the $\alpha^\prime$-expansion of string amplitudes.

CF is supported  by FCT - Fundacao para a Ciencia e Tecnologia, I.P. (2023.01221.BD and DOI  https://doi.org/10.54499/2023.01221.BD). 
HE is supported in part by Department of Energy grant DE-SC0007859.  
JB was supported in part by the Cottrell SEED Award number 
CS-SEED-2023-004 from the Research Corporation for Science Advancement. The work was also supported through computational resources and services provided by Advanced Research Computing at the University of Michigan, Ann Arbor.

JB is grateful to the KITP for hosting him as a KITP Graduate Fellow January - June, 2025. Through this connection, this research was supported in part by the Heising-Simons Foundation, the Simons Foundation, grants no.~NSF PHY-2309135 to the Kavli Institute for Theoretical Physics (KITP). 

HE and JB would like to thank the Niels Bohr International Academy at the Niels Bohr Institute in Denmark for hospitality. 
\appendix
\newpage
\section{Fixing Local Terms at 5 and 6 Points}
\label{app:LocalTerms}

In this appendix, we discuss 5- and 6-point contact terms and to which extent they are fixed by the 5- and 6-point split conditions. The results for the local terms are discussed in Sections \ref{app:5ptsplit} and \ref{app:6ptsplit}. We  connect the structure found with the low-energy expansion of the string tree amplitude in Section \ref{app:MVZ}.

\subsection{5 Point}
\label{app:5ptsplit}

Imposing the split conditions \eqref{eq:5ptSplitA}-\eqref{eq:5ptSplitB} fix all contact terms in the 5-point amplitude up to and including order $X^4$, i.e.~the $w_{i,j}$ coefficients in \reef{eq:5ptLoc} are completely fixed in terms of the 4-point Wilson coefficients $a_{k,q}$ for $i \le 4$. Starting at order $X^5$, some 5-point contact terms have unfixed coefficients. This is summarized in Table \ref{tab:LocalTerms} there the column labelled ``\#locals'' lists the number of independent cyclic polynomials of degree $n$ and the column ``5pt Splits'' shows how many coefficients are left unfixed after imposing \eqref{eq:5ptSplitA}-\eqref{eq:5ptSplitB}. 

It is easy to see why there should be an unfixed contact term at order $X^5$. As explained in Section \ref{s:HZreview}, the $5$-point split happens when we set a dot product of \textit{non-adjacent} momenta to zero; we call these {\em non-planar invariants}. For example, setting $s_{13}=0$ leads to \eqref{eq:5ptSplitA}, and $s_{14}=0$ gives \eqref{eq:5ptSplitB}, and we have similar (cyclically related) splits when we set $s_{24}$, $s_{25}$, or $s_{35}$ to zero. These non-planar invariants can be expressed in terms of the planar ones, $X_{i,j}$, as follows:
\begin{equation}
    s_{ij} = 2 p_{i} \cdot p_{j} = -X_{i,j} - X_{i+1,j+1} +X_{i,j+1}+X_{i+1,j}\,,
\end{equation}
The degree 5 polynomial obtained as the product of the five non-planar invariants,
\begin{equation}
P_{X^5}^{\text{unfixed}} = s_{13} s_{14} s_{24} s_{25} s_{35}\,,
\end{equation}
vanishes on all five different split loci and therefore no constraints are imposed on the coefficient of this polynomial. This accounts for the single free 5-point coefficient in Table \ref{tab:LocalTerms}. 
Moreover, at higher orders, starting at $X^6$, the free terms can be obtained by multiplying this polynomial by any other cyclically symmetric polynomials. The numbers in Table \ref{tab:LocalTerms} exactly reproduce the count of cyclic polynomials in the first column. It is perhaps a bit surprising that these are the only ones (at least up to order $X^{13}$ computed). 

\begin{table}[t]
\centering
\begin{tabular}{c|ccc|cc}
\multirow{2}{*}{$\mathcal{O}(X^n)$} & \multicolumn{3}{c|}{\textbf{5-point}}                                                & \multicolumn{2}{c}{\textbf{6-point}}               \\
                      & \multicolumn{1}{c|}{\#locals} & \multicolumn{1}{c|}{5pt Splits} & 6pt Splits & \multicolumn{1}{c|}{\#locals} & 6pt Splits \\ \hline
0                     & 1                             & 0                               & 0          & 1                             & 0          \\
1                     & 1                             & 0                               & 0          & 2                             & 0          \\
2                     & 3                             & 0                               & 0          & 9                             & 0          \\
3                     & 7                             & 0                               & 0          & 32                            & 0          \\
4                     & 14                            & 0                               & 0          & 89                            & 0          \\
5                     & 26                            & 1                               & 1          & 226                           & 0          \\
6                     & 42                            & 1                               & 1          & 523                           & 2          \\
7                     & 66                            & 3                               & 2          & 1104                          & 9          \\
8                     & 99                            & 7                               & 3          & 2194                          & -          \\
9                     & 143                           & 14                              & -          & -                             & -          \\
10                    & 201                           & 26                              & -          & -                             & -          \\
11                    & 273                           & 42                              & -          & -                             & -          \\
12                    & 364                           & 66                              & -          & -                             & -          \\
13                    & 476                           & 99                              & -          & -                             & -         
\end{tabular}
\caption{\label{tab:LocalTerms}Counting of local terms at $5$-points and $6$-points, and how these get fixed by the $5$- and $6$-point split constraints.}
\end{table}

\subsection{6 Point}
\label{app:6ptsplit}

Let us now continue this analysis by understanding what happens when we include the split constraints coming from the $6$-point amplitude. There are two types of splits at 6 point: one where the $6$-point amplitude factors into a product of three $4$-point amplitudes, and another where it factors into a $5$-point times a $4$-point. The first one is a simple consequence of the second one, provided that the $5$-point amplitude also obeys the $5$-point splitting conditions. For this reason we focus solely on the second type of split which can be stated as \cite{Arkani-Hamed:2023swr}
\begin{equation}
\begin{aligned}    &g\mathcal{A}_6\big(s_{13}=s_{14}=0\big) \,=\,   \mathcal{A}_4\big(X_{1,3},X_{2,6}\big) \times \mathcal{A}_5\big(X_{2,4},X_{2,5},X_{3,5},X_{3,6},X_{4,6}\big)\,,\\
&g\mathcal{A}_6\big(s_{13}=s_{15}=0\big) \,=\,  \mathcal{A}_4\big(X_{1,3},X_{2,6}\big) \times \mathcal{A}_5\big(X_{2,4},X_{1,5},X_{3,5},X_{3,6},X_{4,6}\big)\,,\\
&g\mathcal{A}_6\big(s_{14}=s_{15}=0\big) \,=\,   \mathcal{A}_4\big(X_{1,3},X_{2,6}\big) \times \mathcal{A}_5\big(X_{1,4},X_{1,5},X_{3,5},X_{3,6},X_{4,6}\big)\,.\\
\end{aligned}
\label{eq:6ptSplits}
\end{equation}
The respective cyclic permutations of these are automatically enforced since we consider a cyclically symmetric $6$-point ansatz. To implement  the constraints \reef{eq:6ptSplits} we construct the most general cyclically symmetric ansatz for the $6$-point EFT amplitude, just like we did in Section \ref{s:HZEFT} for the $5$-point one. At $6$ points, we divide $\mathcal{A}_6$ into the following parts: 
\begin{itemize}
\itemsep-0.1em
    \item $\mathcal{A}^{(0)}_6$, which contains three poles, corresponding to purely cubic diagrams; 
    \item $\mathcal{A}^{(1)}_6$, which contains two poles, corresponding to diagrams with two cubic and a quartic interaction;
    \item $\mathcal{A}^{(2)}_6$, which contains a single pole, corresponding to two types of diagrams: those with a cubic vertex and a $5$-point contact term or those with two quartic contact terms;
    \item  $\mathcal{A}^C_6$, which is the purely local part, parametrized by all possible cyclic polynomials. (At degree $k$, we use $z_{k,q}$ to denote coefficients of the independent cyclic polynomials.)
\end{itemize}  
The dependence of $\mathcal{A}_6$ on the $4$-point Wilson coefficients $a_{k,q}$ is introduced via both $\mathcal{A}^{(1)}_6$ and $\mathcal{A}^{(2)}_6$, while the $5$-point Wilson coefficients, $w_{i,j}$ enter exclusively via $\mathcal{A}^{(2)}_6$.

The results are shown in the right-hand columns of Table \ref{tab:LocalTerms}. Even though the number of locals terms at $6$-points grows exponentially, by imposing the $6$-points splits \eqref{eq:6ptSplits}, we are able to fix all local terms up to order $X^6$, where just $2$ coefficients are unfixed out of a total of $523$. 

Notably, the 6-point splitting conditions also put constraints on the 5-point coefficients $w_{i,j}$. We first see this at order $X^7$ in the 5-point amplitude where 3 coefficients were left free after imposing the 5-point splits, but one of these gets fixed by the 6-point splits, leaving only two coefficients free. Similarly, at order $X^8$, 7 free coefficients are reduced to just 3. One may anticipate that going to higher order --- and incorporating higher-point splits  --- would fix additional free coefficients. 

\subsection{Relation to the String Amplitude}
\label{app:MVZ}

It is well-known that the coefficients of the different orders in the $\alpha^\prime$-expansion of string amplitudes are given by \textit{multiple-zeta values} (MZVs) \cite{Schlotterer:2012ny,Brown:2013gia,Broedel:2013aza,Blumlein:2009cf}. At a given order $(\alpha^\prime)^n$, we can only find MZVs of weight equal to $n$ (or products of MZVs of lower weight, but such that the total weight is $n$). While the $\alpha^\prime$-expansion at $4$-point only contains \textit{single} zeta values, $\zeta_i$, at higher points, we start seeing the appearance of MZVs with other depths. Concretely, we anticipate finding the first linearly independent (or primitive) MZV, $\zeta_{3,5}$, at order $(\alpha^\prime)^8$, and the next ones, $\zeta_{3,7}$ and $\zeta_{3,3,5}$, at orders $(\alpha^\prime)^{10}$ and $(\alpha^\prime)^{11}$, respectively. From this reasoning, we cannot expect to fix \textit{all} $5$-point Wilson coefficients in terms of $4$-point ones. Otherwise, when applied to the string amplitude, this would mean that we could write primitive higher-depth MZVs in terms of rational linear combinations of single zeta values (with the same total weight) entering in the $4$-point $a_{k,q}$'s -- which is widely believed to be impossible since primitive MZVs are conjectured to be linearly independent over the rational numbers. However note that it is not the case that all allowed MZVs at a given order in $\alpha^\prime$,  have to appear in the expansion of the amplitude.

For example, as mentioned, the MZV $\zeta_{3,5}$ could appear on its own at order $X^6$ in the $5$-point amplitude. Therefore, if $\zeta_{3,5}$ appears in the $\alpha^\prime$ expansion of the $5$-point amplitude, then it must multiply precisely the polynomial left unfixed at this order. However, using \cite{Mafra:2022wml} we can check that, surprisingly, for the diagonal $Z$-theory integral under consideration, $\zeta_{3,5}$ does \textit{not} appear in the $5$-point $\alpha^\prime$ expansion. This means that the free coefficient at order $X^6$ is also a simple zeta value.\footnote{Note that the vanishing of $\zeta_{3,5}$ for the $5$-point diagonal $Z$-theory integral does not mean that no higher-depth MZVs appear in the $\alpha^\prime$ expansion of this amplitude. In particular one can explicitly check that $\zeta_{3,3,5}$ does appear. Of course, $\zeta_{3,5}$ might also still appear multiplied by other $\zeta$-values at higher orders in the expansion. However, for the purpose of our analysis, we are focusing on the MZVs that are relevant at the orders in $\alpha^\prime$ that we can significantly constrain as shown in Table \ref{tab:LocalTerms}.}

At $6$-point, the first primitive MZV, $\zeta_{3,5}$, is expected to appear at order $X^5$. At this order, we find that there are no free $6$-point local terms after imposing \eqref{eq:6ptSplits}, and so all get fixed in terms of $5$- and $4$-point local terms. Therefore, we must have that $\zeta_{3,5}$ (on its own) also does not appear in the $6$-point amplitude. Indeed using \cite{Mafra:2022wml}, we can check that this is precisely the case. So without actually performing the $\alpha^\prime$ expansion, and simply analysing the number of free local terms post splits, we can directly infer about the presence/absence of certain  MZVs and, if present, predict the polynomial in $X$ they multiply.

A natural question is then whether imposing the constraints from splits of higher-point amplitudes fully fixes all  local terms, apart from those given by MZV's for the string and that therefore cannot be decomposed in terms of $4$-point Wilson coefficients. As we increase $n$, the number of local terms grows exponentially, but at the same time there are more and more split patterns to impose, so there's a hope that this could indeed be true. We leave this discussion for future work.

\section{Implementation of nonlinear constraints}
\label{app:nonlin}
We outlined in Section \ref{s:NonLin} how we linearized the nonlinear constraints in order to implement them in SDPB. Here we provide further details about the constraints and how the double-scans are done for the island plots in the main text. 

\subsection{Constraints in terms of $X$ and $Y$}
When equation \reef{gSQ} is used to eliminate $g^2$ from the nonlinear  constraints in Table \ref{tab:Constraintsk6}, we obtained a set of $\lfloor k/2 \rfloor$ nonlinear constraints for each $k\ge 3$. Replacing $a_{1,0}$ and $a_{2,1}$ in these constraints in favor of $X$ and $Y$ defined as in  \reef{defXY} as
\begin{equation}
\label{defXY2}
    X = \frac{a_{1,0}}{a_{0,0}} 
    \,, ~~~~~~~
    Y = \frac{3a_{2,0}-2a_{2,1}}{a_{0,0}}
     \, ,
\end{equation}
we get 
\be  \label{nonlinA}
\begin{array}{llrcl}
   k=3\!: &~~~~& 
   X Y a_{0,0}
   -
   2a_{3, 0} - a_{3, 1} &=& 0 \,,  
 \\[2mm]
   k=4\!: &~~~~& 
   6 X^2 Y a_{0,0} - Y^2 a_{0,0} + 9 Y a_{2,0} - 20 a_{4,0} + 
 6 a_{4,2}&=& 0
   \,, \\[1mm]
    &~~~~& 
   X^2 Y a_{0,0} + 2 Y a_{2,0} 
   - 5 a_{4, 0} + 2 a_{4,1} &=& 0
 \,,
 \\[2mm]
 k=5\!: &~~~~& 
   -X Y^2 a_{0,0} + 5 X Y a_{2,0} 
   + 4 Y a_{3,0} - 10 a_{5,0} 
   + 2 a_{5,2}&=& 0\,, 
   \\[1mm]
    &~~~~& 
   X Y a_{2,0} + Y a_{3,0} - 3 a_{5,0} + a_{5,1} &=& 0 \,,
   \\[2mm]
   k=6\!: &~~~~& 
   -24 X^2 Y^2 a_{0,0} 
   + 51 Y \frac{a_{2,0}^2}{a_{0,0}}
   + Y^3 a_{0,0} 
   - 18 Y^2 a_{2,0}
   \phantom{space}  \\
   &~~~~&
   + 96 X Y a_{3,0} 
   + 80 Y a_{4,0} 
   - 210 a_{6,0} + 24 a_{6,3}&=& 0\,, 
   \\[1mm]
    &~~~~& 
   -X^2 Y^2 a_{0,0} 
   + 3 Y \frac{a_{2,0}^2}{a_{0, 0}}
   - Y^2 a_{2,0}  
   + 6 X Y a_{3,0} 
   \phantom{space}  \\
   &~~~~& 
   + 5 Y a_{4,0} 
   - 14 a_{6,0} 
   + 2 a_{6,2}
 &=& 0 \,,
    \\[1mm]
    &~~~~& 
   Y \frac{a_{2,0}^2}{a_{0, 0}} 
   + 2 X Y a_{3,0} 
   + 2 Y a_{4,0} 
   - 7 a_{6,0} 
   + 2 a_{6,1}
    &=& 0 \,.
   \\[2mm]
   \end{array}
\ee
For the double scan over both $X$ and $Y$, we directly use these relations for $k \le 5$. However, at $k=6$ the $a_{2,0}^2$ terms obstruct the linearization. Instead of additionally scanning over $a_{2,0}/a_{0,0}$, we eliminate them by solving for $a_{2,0}^2$ in the last $k=6$ constraint in \reef{nonlinA}. Then we can  impose two of the three $k=6$ nonlinear constraints in the double scan, namely 
\be  \label{nonlinAk6XY}
\begin{array}{llrcl}
   k=6\!: &~~~~&
   24 X^2 Y^2 a_{0,0} 
   + Y^3 a_{0,0} 
   - 18 Y^2 a_{2,0} 
   - 6 X Y a_{3,0} 
   - 22 Y a_{4,0} 
   \phantom{space}\\
   &~~~~&
   + 147 a_{6,0} 
   - 102 a_{6,1} 
   + 24 a_{6,3}
 &=& 0 \,,  
 \\[1mm]
 &~~~~&
 -X^2 Y^2 a_{0,0} 
 - Y^2 a_{2,0} 
 - Y a_{4,0} 
 + 7 a_{6,0} 
 - 6 a_{6,1} 
 + 2 a_{6,2}
 &=& 0 \,.
   \end{array}
\ee
Likewise, at $k=7$ we can include two out of the three nonlinear constraint when we only scan over $X$ and $Y$, while at $k=8$ it is two out of four. 

To impose nonlinear constraints with the single scan over just $Y$, we use the second $k=4$ equation to eliminate each  appearance of $X^2$. This leaves one constraint at $k=4$ and the two from \reef{nonlinAk6XY} at $k=6$. At  $k=5$, we can only use the linear combination of the two constraints without $X Y a_{2,0}$. 
To summarize, for $k \le 8$, we have the following $Y$-scan constraints
\be  \label{nonlinAk8Yonly}
\begin{array}{llrcl}
   k=3\!: &~~~~&
   Y\, a_{1,0}
   = 
   2a_{3, 0} - a_{3, 1} &=& 0\,,
   \\[2mm]
   k=4\!: &~~~~&
   - Y^2 a_{0,0} - 3 Y a_{2,0} 
   + 10 a_{4,0} 
   - 12 a_{4,1}
   + 6 a_{4,2} 
    &=& 0 \,,
    \\[2mm]
   k=5\!: &~~~~&
   -Y^2 a_{1,0} 
   - Y a_{3,0} 
   + 5 a_{5,0} 
   - 5 a_{5,1} 
   + 2 a_{5,2}
   &=& 0 \,,
   \\[2mm]
   k=6\!: &~~~~&
    Y^2 a_{2,0} 
    - 6 Y a_{4,0} 
    + 2 Y a_{4,1} 
    + 7 a_{6,0} 
    - 6 a_{6,1} 
    + 2 a_{6,2}
    &=& 0 \,,
    \\[2mm]
    k=7\!: &~~~~&
    \text{none}
    \\[2mm]
    k=8\!: &~~~~&
    Y^4 a_{0,0} 
    + 15 Y^3 a_{2,0} 
    - 100 Y^2 a_{4,0} 
    + 30 Y^2 a_{4,1} 
    + 300 Y a_{6,0} 
    \phantom{space}\\
   &~~~~&
    - 150 Y a_{6,1}    
    + 30 Y a_{6,3} 
    - 306 a_{8,0} 
    + 300 a_{8,1} 
    - 150 a_{8,2} 
    + 30 a_{8,4}
    &=& 0 \,.
   \end{array}
\ee
These are the linearized splitting constraints we imposed when we compute the $Y$-scan bounds in the main text. 

\subsection{Numerical Implimentation of the Double-Scans}
\begin{figure}[t]
    \centering
    \includegraphics[width=\linewidth]{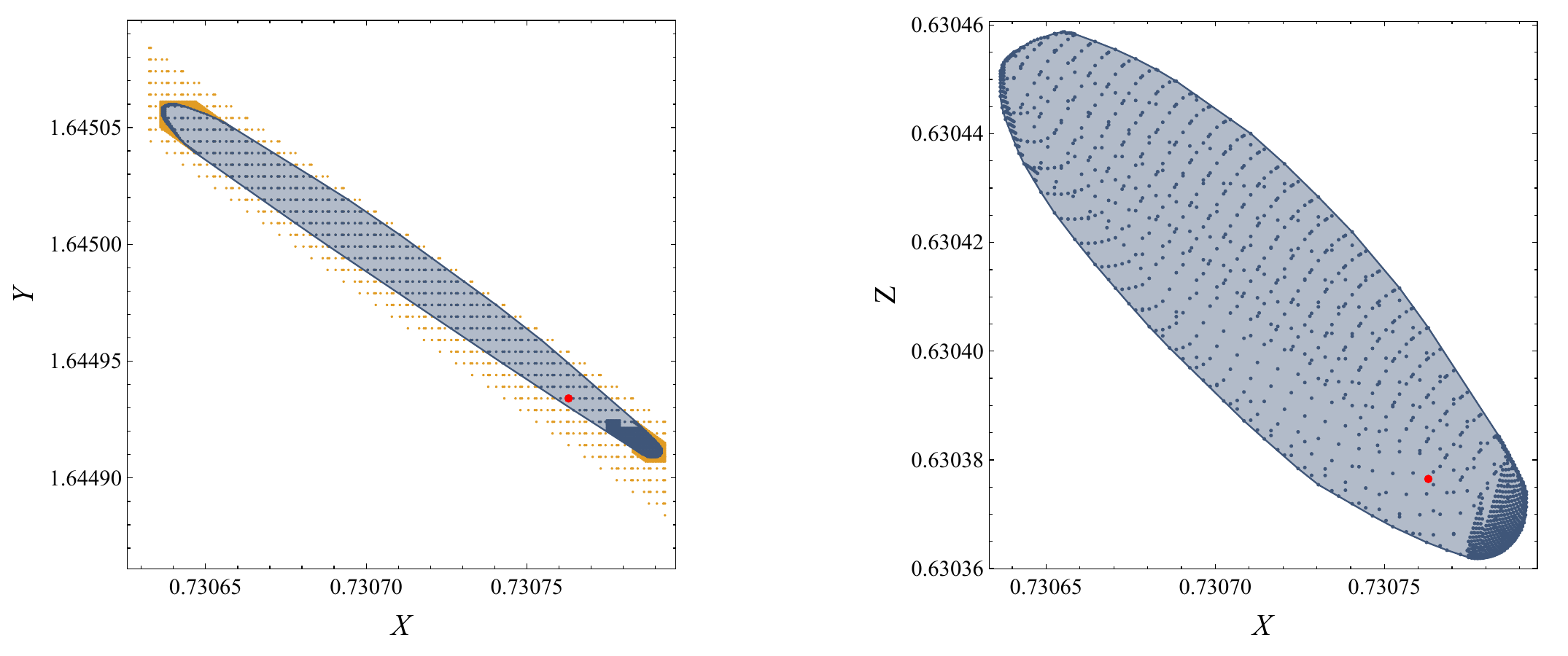}
    \caption{ 
    {\bf Left:} Feasibility scan over points $(X,Y)$ to determine which are allowed when imposing the eleven $k\le 8$ nonlinear constraints that can be linearized for fixed values of $X$ and $Y$. Orange points show sample points that did not pass, blue are points that passed (i.e.~for which SDPB found a solution).  
    {\bf Right:} For each blue point in the $(X,Y)$-plane on the left, we compute the maximum and minimum of $Z = a_{3,0}/a_{0,0}$. The projection of the allowed $(X,Y,Z)$ region to the $(X,Z)$ plane is shown on the right.}
    \label{fig:kmax8scans}
\end{figure}

When scanning over possible values of $X$ and $Y$, we can impose seven nonlinear conditions at $k_{\max}=6$ and eleven at $k_{\max}=8$. The resulting allowed region in the $(X,Y)$-plane becomes a {\em feasibility region} in that it selects the points in $(X,Y)$ for which SDPB converges to a solution and excludes points where it fails to find a solution.\footnote{This can be quite sensitive to the choice of spins included in the finite truncation of the sum over spins in the dispersion relations, so we make sure to first benchmark spinlist choices before the feasibility runs are done.} 
As an example, the plot on the left in Figure \ref{fig:kmax8scans} shows the sample points (orange) and the passed points (blue) for the $k_{\max} = 8$ double-scan island. It is not a given that the islands should be convex, nonetheless that is what we have found in each case. 
The dark blue island shown in Figure
\ref{fig:Zoom10d} (and its zoomed-in version on the left of Figure \ref{fig:kmax8islandszoom}) was obtained as the convex hull of the blue points in Figure \ref{fig:kmax8scans}.  

The interior points in the island are not needed to determine its shape. However, they are needed to determine the full range of $Z= a_{3,0}/a_{0,0}$, as needed for the $(X,Z)$ island. The plot on the right in Figure \ref{fig:kmax8scans} shows the resulting points in the $(X,Z)$ when the max and min of $Z$ are computed for each of the feasible points on in the $(X,Y)$ island. This is how we obtain the  $(X,Z)$ in Figure \ref{fig:kmax8islandszoom}. 
The method of scanning over points is rather primitive, but sufficient to illustrate the impact of the nonlinear constraints. 

\section{Results From Twice-Subtracted Dispersion Relations}
\label{app:2sdr}
In Section \ref{sec:intro}, we discussed the assumed behavior of the amplitude in the Regge limit (large $s$ fixed momentum transfer $u$) as being such that
\be
   \label{impRegge}
    \lim_{|s|\to\infty} \frac{\mathcal{A}_4(s,u)}{s} \to 0 \, .
\ee
This assumption, while harder than the Regge behavior of the actual string amplitude, which itself vanishes in the Regge limit, was still enough to uniquely pick out the string amplitude under the assumption of finite spin. However, the Regge behavior \reef{impRegge} is not generic. The Froissart-Martin bound \cite{Froissart:1961ux, Martin:1962rt} instead suggests that amplitudes are required to satisfy only
\be
   \label{FroisRegge}
    \lim_{|s|\to\infty} \frac{\mathcal{A}_4(s,u)}{s^2} \to 0\, ,
\ee
leading to ``twice-subtracted'' dispersion relations (twice-subtracted because one has two subtraction terms, one at order $\mathcal{O}(s^0)$ and the other at order $\mathcal{O}(s)$).

In this appendix we discuss the impact of relaxing the Regge assumption \reef{impRegge} to \reef{FroisRegge}. The dispersion relations for $c_{k,k-1}$ Wilson coefficients are then no longer convergent and therefore the fixed-$u$ crossing conditions
\begin{align}
    c_{k,q}- c_{k,k-q} = 0
\end{align}
can only be applied for $2\leq q \leq k-2$. Similarly, in the fixed-$t$ constraints, we now eliminate both $b_{k,k}$ and $b_{k,k-1}$ from the null constraints \reef{asfrombs}, so that \eqref{asfrombs1sdr} is replaced with
\begin{align} \label{asfrombs2sdr}
    c_{k,q}- \binom{k}{q}c_{k,0} +\binom{k-1}{q}(k~c_{k,0}-c_{k,1}) = \sum_{q'=q}^{k-2}(-1)^{q'}\binom{q'}{q}b_{k,q'} \, .
\end{align}
Additionally, the splitting condition
\begin{align}
    g^2 = \frac{a_{0,0}}{2a_{2,1}-3a_{2,0}}\,
\end{align}
can no longer be used directly, since $a_{0,0} = c_{1,0}$ does not have a convergent dispersion relation. Instead, we must use both the $k = 2$ and $k = 3$ conditions to solve for both $g^2$ and $a_{0,0}$:
\be
    g^2 = \frac{a_{1,0}^2(3a_{2,0}-2a_{2,1})}{(2a_{3,0}-a_{3,1})^2}\,,
    ~~~~~~
    a_{0,0} = \frac{a_{1,0}(3a_{2,0}-2a_{2,1})}{2a_{3,0}-a_{3,1}}\, .
\ee
It is then convenient to define $X'$ and $Y'$ such that
\be
    X' = \frac{3a_{2,0}-2a_{2,1}}{a_{1,0}}\,,
    ~~~~~~~~~
    Y' = \frac{2a_{3,0}-a_{3,1}}{a_{1,0}}\,,
\ee
which allow us to linearize the split constraints starting at $k = 4$. 

\begin{figure}[t]
    \centering
    \includegraphics[width=0.9\linewidth]{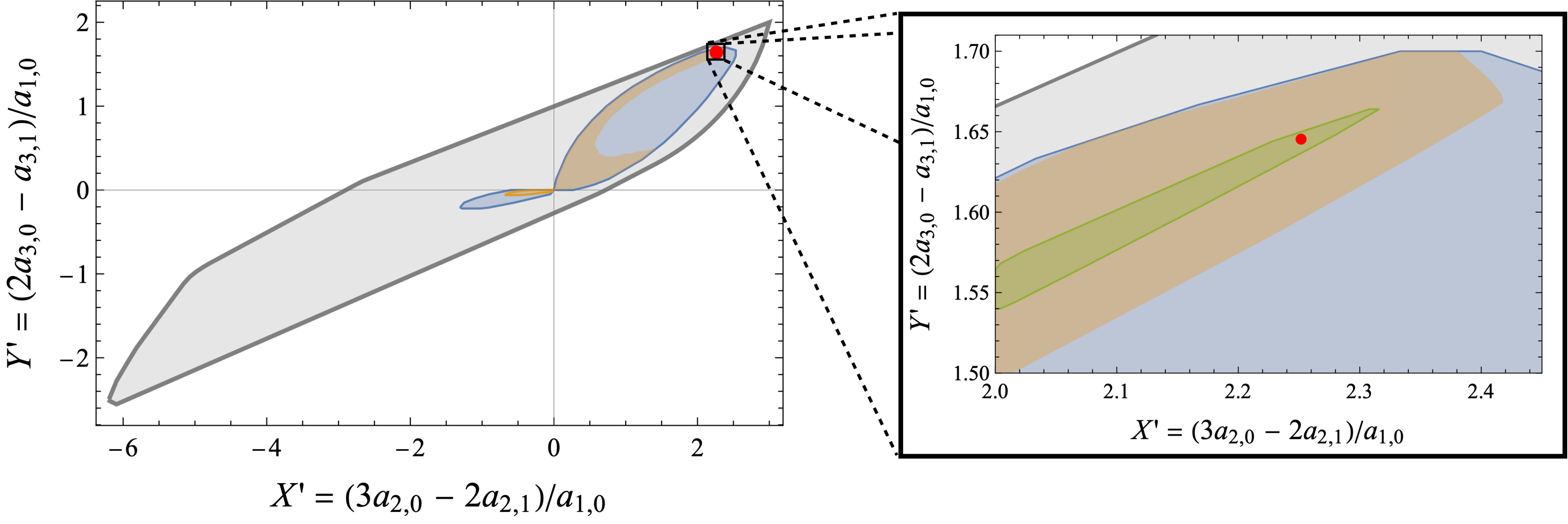}
    \caption{\textbf{Left:} The $(X,Y)$ with the more generic Regge behavior at $k_{\max} = 4$ in 10D. The gray region is allowed when the splitting constraints are not imposed, while the blue region is allowed with the splitting constraints but without spectrum input. The orange region shows the $k_{\max} = 4$ allowed region with a scalar, vector, and spin-two state allowed at the mass gap and no other states until $\mu_c = 2$.\newline
    \textbf{Right:} A zoom in near the tip of the orange allowed region with spectrum input. In green we show the much thinner allowed region for $k_{\max} = 5$.    
    }
    \label{fig:2sdrXY}
\end{figure}

In Figure \ref{fig:2sdrXY}, we show the $k_{\max} = 4$ allowed region in $d=10$ without the splitting constraints (gray) and  with splitting constraints but no spectrum input (blue). With both splitting constraints and allowing only a scalar, vector, and spin-two state at  $M_\text{gap}^2$ and no other states up to cutoff $\mu_c = 2$, we get the orange region; as shown in \cite{Berman:2024kdh,Berman:2024owc}, $j=0,1,2$ are only spins at the mass gap in the twice-subtracted bootstrap when assuming a finite spectrum at $M_\text{gap}$ and $\mu_c > 1$. 
As in the case of low $k_{\max}$ in the $d = 4$ and $\mu_c = 1.2$ bootstrap discussed in Section \ref{s:D4islands},  the region with spectrum input shown in Figure \ref{fig:2sdrXY} does not bifurcate at $k_\text{max}=4$. There is also a  small region of allowed parameter space for $X',Y' < 0$ at finite $k_{\max}$, but we expect that this region shrinks as we increase $k_{\max}$.

In Section \ref{s:D4islands}, we saw that increasing the number of constraints lead to bifurcation: at large enough $k_{\max}$, the allowed region became disconnected with an island around the string amplitude. Here, however, as we increase $k_{\max}$, the allowed region shrinks, but we cannot definitively show that it bifurcates. In particular, because we always have to do a double scan over $X'$ and $Y'$ to linearize the constraints, there is always a chance that when we run a grid of points, some thin allowed region escapes out of that grid. This was not a concern in the cases discussed in Sections \ref{s:D10islands} and \ref{s:D4islands}, because we found that the region bifurcated with just the $Y$-scan constraints \reef{nonlinAk8Yonly} and because the double scan islands had reasonably convex shapes whose boundaries we could convincingly determine. As we increase $k_{\max}$ for this more general scenario, the ``bridge'' region connecting the string amplitude to the scaled down bulk region continues to thin, but it is not clear whether it actually pinches off into two separate regions. An example of this is shown on the left of Figure \ref{fig:2sdrXY}, with the $k_{\max} = 5$ allowed region with spectrum input displayed in green. The allowed region thins, but does not appear to break anywhere. This behavior persists and, even in tests at $k_{\max} = 8$, there are feasible solutions for $(X',Y')$ with $Y' <\zeta_2-1/10$ and for $X$ in a range  only $1/1000$th wide. This means  that, even if there were to be an island, it would be at least two orders of magnitude longer than it is wide. Therefore, an extremely fine scan would be necessary to even somewhat convincingly show whether the region bifurcates or not. 

It would be interesting to use some of the boundary finding methods developed for the conformal bootstrap, for example the Skydiving algorithm \cite{Liu:2023elz}, to search more systematically for a boundary of the island, but the region may not bifurcate, or at  least it may take the includsion of a larger set of nonlinear constraints to see it.

\bibliography{main}

\end{document}